\documentclass[a4paper,11pt]{article}
\usepackage{times}
\usepackage{array}
\usepackage{amsmath}
\usepackage{latexsym}
\usepackage{amsfonts}
\usepackage{pifont}
\usepackage{stmaryrd}
\usepackage{amssymb}
\usepackage{array}
\usepackage{epsfig}
\usepackage[labelformat=default, labelsep=default, justification=centerlast,  font=footnotesize]{caption}
\usepackage{verbatim}
\usepackage{multirow}
\usepackage{rotating}
\usepackage{amsthm}
\usepackage{enumerate}
\usepackage{color}
\usepackage[dvipsnames]{xcolor}
\usepackage[right=2.75cm, left=2.75cm, top=2.75cm, bottom=3cm]{geometry}

\usepackage{comment}
\usepackage[colorlinks]{hyperref}
\usepackage{sidecap}
\usepackage{graphicx}

\usepackage[normalem]{ulem} 
\usepackage{cancel} 

\newcommand\id{\leavevmode\hbox{\small1\kern-3.3pt\normalsize1}}

\def\f12{\frac{1}{2}}

\newcommand{\tr}{\mbox{Tr}}

\newcommand{\Bra}[1]{{ \langle \! \langle{#1}\vert }}
\newcommand{\Ket}[1]{{ \vert {#1}  \rangle \!  \rangle}}
\newtheorem*{definition*}{Definition}

\def\go{g_\mathrm{00}}

\newcommand{\bra}{\langle}
\newcommand{\ket}{\rangle}

\newcommand{\be}{\begin{equation}}
\newcommand{\ee}[1]{\label{#1} \end{equation}}
\newcommand{\bbe}{\begin{equation*}}
\newcommand{\eee}{\end{equation*}}

\newcommand{\bs}{\begin{split}}
\newcommand{\es}{\end{split}}

\def\f12{\frac{1}{2}}

\title{Bell's Theorem for Temporal Order}
\author{Magdalena Zych,$^{1\ast}$ Fabio Costa,$^{1}$ Igor Pikovski,$^{2,3,4}$ \v{C}aslav Brukner$^{5,6}$
\\
\normalsize{$^{1}$ Centre for Engineered Quantum Systems, School of Mathematics and Physics,}\\ 
\normalsize{The University of Queensland, St Lucia, QLD 4072, Australia }\\
\normalsize{$^{2}$ITAMP, Harvard-Smithsonian Center for Astrophysics, Cambridge, MA 02138, USA}\\
\normalsize{$^{3}$Department of Physics, Harvard University, Cambridge, MA 02138, USA}\\
\normalsize{$^{4}$Stevens Institute of Technology, Hoboken, NJ 07030 USA}\\
\normalsize{$^{5}$ Vienna Center for Quantum Science and Technology (VCQ),}\\
\normalsize{ University of Vienna, Faculty of Physics, Boltzmanngasse 5, A-1090 Vienna, Austria}\\
 \normalsize{$^{6}$Institute for Quantum Optics and Quantum Information (IQOQI),}\\
\normalsize{ Austrian Academy of Sciences, Boltzmanngasse 3, A-1090 Vienna, Austria}\\
\normalsize{$^\ast$ E-mail: m.zych@uq.edu.au}
}
\date{}
\begin{document}
\maketitle

\begin{quote}
Time has a fundamentally different character in quantum mechanics and in general relativity. In quantum theory events unfold in a fixed order while in general relativity temporal order  is influenced by the distribution of matter. When matter requires a quantum description, temporal order is expected to become non-classical -- a scenario beyond the scope of current theories.  Here we provide a direct description of such a scenario. We consider a thought experiment with a massive body in a spatial superposition and show how it leads to entanglement of temporal orders between time-like events.
 This entanglement enables accomplishing a task, violation of a Bell inequality, that is impossible under local classical temporal order; it means that temporal order cannot be described by any pre-defined local variables. A classical notion of a causal structure is therefore  untenable in any framework compatible with the basic principles of quantum mechanics and classical general relativity.
\end{quote}

\section*{Introduction}

Quantum mechanics forces us to question the view that physical quantities (such as spin, positions or energy) have predefined values:
Bell's  theorem shows that if observable quantities were determined by some locally-defined {classical} variables, it would be impossible to accomplish certain tasks -- such as the violation of Bell's inequalities -- whereas such tasks are possible according to quantum mechanics~\cite{bell64, Clauser1969} {and have been realised in experiments~\cite{Clauser:1972, hensen2015experimental, Giustina2015, Shalm2015}}. However, the {causal relations} between events remain fixed in quantum theory: whether an event $\mathrm A$ is in the past, in the future, or space-like separated from another event $\mathrm B$ is pre-defined by the location of such events in space-time~\cite{Hardy:2005, Hardy:2007bk}. In contrast, in general relativity, space-time itself is dynamical: the presence of massive objects affects local clocks and thus 
causal relations between events defined with respect to them. 
Nonetheless, the dynamical causal structure of general relativity is still classically predefined: the causal relation between any pair of events is uniquely determined by the distribution of matter-energy degrees of freedom in their past light-cone. In other words, causal relations are always determined by local classical variables. The picture is expected to change if we consider quantum states of gravitating degrees of freedom: if a massive system is prepared in a superposition of two distinct states, each yielding an observably different causal structure for future events, would it be possible to observe causal relations which display genuine quantum features?  

A main obstacle in the analysis of macroscopic superpositions of gravitating bodies is that, in the absence of a classical space-time manifold, it becomes unclear how to identify space-like surfaces on which quantum states live, or global fields of time-like vectors to define time evolution. Indeed, some models even postulate that such superpositions are simply not valid physical states and must decohere (or collapse) fast enough to preserve a classical description of space-time and dynamical laws~\cite{ref:Karolyhazy1966, ref:Diosi1989, ref:Penrose1996, Stamp:2012, Penrose2014}.
A very different mindset underlies various quantum gravity frameworks~\cite{kiefer2012quantum} -- where quantum features of the metric and therefore of the causal relations are indeed expected. However to date none of the quantum gravity frameworks has been applied to analyse such an epitomic example as  superpositions of space-times with macroscopically distinct causal structures. Therefore, it is unclear whether there exists any phenomenology unequivocally associated with quantum causal structures, nor whether quantum gravity frameworks can  circumvent or directly  address the objections against superpositions of manifolds that motivate the collapse models.
Independently, quantum formalisms have been recently developed to study quantum causal structures at an abstract level in the context of quantum information processing~\cite{Hardy:2007bk, Chiribella:2013, Oreshkov:2012}. However, although quantum features of space-time are among the motivations for these studies, no direct link with quantum gravity has yet been established.

This work provides the first direct analysis of quantum causal relations arising from a spatial superposition of a massive object.
We show how the temporal order between time-like events can become superposed or even entangled. We further discuss a thought experiment, an admissible albeit remote physical scenario, where these non-classical causal relations arise among physical events. In order to prove their non-classicality, we formulate a Bell-type theorem for temporal order: We define a task that cannot be accomplished if the time order between the events was predetermined by  local variables, while the task becomes possible if the events are in a space-time region affected by the gravitational field of a massive object in an appropriate quantum state.
Our approach provides a method to directly describe scenarios so far considered to be out of reach for standard theoretical physics. 
We show explicitly how to overcome the difficulties with describing superpositions of metrics that motivated collapse models. On the other hand, our result is independent of the high-energy completion of any specific quantum gravity framework --  we do not assume any new physics, basing entirely on well-established, low-energy general relativity and on quantum mechanics. Our results are therefore robust against particular mathematical approaches to quantising gravity, thus providing a benchmark for specific frameworks. Furthermore, the time and energy scale at which entangled temporal order arises is far above the Planck scale, typically invoked in this context, and is also far remote from the scale given by the decoherence models -- which therefore do not preclude quantum features of space-time to arise. Our results thus reveal that both the above approaches are missing crucial intuition and correct physical understanding of the phenomena associated with causal structures at the interface of quantum and gravitational physics. In turn, our work provides  a robust method to quantitatively assess these phenomena and build correct physical intuition for quantum causal structures.

\section*{Results}
\subsection*{Dynamical causal structure in general relativity}
\label{sec:time_dilation}

In classical general relativity, the causal structure is the structure of light cones of the spacetime metric~\cite{Hawking1976, Malament1977}. As the matter-energy degrees of freedom (DOFs) determine the metric through Einstein's equations, the causal structure of a region of spacetime is dynamical: it depends on the state of the matter-energy in its past light cone.
A major obstacle towards a quantum theory of gravity is that it is not clear how to transpose the mathematical notion of causal relations to scenarios where matter DOFs can be in general quantum states, as such scenarios seem to preclude the use of any underlying space-time manifold with respect to which events, light cones, and causal relations could be defined. To overcome this obstacle, our approach is to start from a physical understanding of events and their causal relations.
Even in classical general relativity a physical event cannot be directly identified with a point on a space-time manifold, a fundamental aspect of the theory captured mathematically by diffeomorphism invariance~\cite{Stachel2014}. Although it can be debated whether or not space-time points have an intrinsic physical meaning, a natural way to define diffeomorphism-invariant  events is to specify them operationally, relative to physical systems, for example: positions and proper times of physical systems used as clocks~\cite{Rovelli1991}. We adopt this notion of events throughout the work. Causal relations are then understood as the possibility to exchange non-faster than light signals---or more generally, physical systems---between operationally defined events.

The presence of massive bodies {generally} alters the {relative} rates at which clocks tick. For example, in a weak field limit, a clock in a gravitational potential $\Phi$ exchanging signals with an identical clock far away from the source of $\Phi$, where the potential effectively vanishes, will appear to tick slower by a factor $\sqrt{1+2 \frac{\Phi}{c^2}}$.
In classical physics, this  leads to the well-tested time dilation~\cite{HafeleKeating:1972b, Wineland:2010} and redshift effects~\cite{PoundRebka:1960}. When the clocks are described as quantum systems, new effects arise from the combination of quantum and general relativistic theories. For a clock in superposition of different distances to the mass, its {time-keeping degrees of freedom become}
entangled to the clock's position~\cite{Zych:2011, Zych:2012, Zych2016}. This entanglement implies a universal decoherence mechanism for generic  macroscopic systems under time dilation~\cite{pikovskiuniversal2015, pikovski2017time}. The regime of low-energy quantum systems in curved space-time can be described within a framework of general-relativistic composite quantum particles \cite{ZychPhDSpringer}. Here we additionally exploit the fact that only the distance between a clock and a mass has physical significance and due to linearity of quantum theory this must hold also for a superposition of different distances. (There is no difference in the relative ticking rates of two clocks whether we think that the clocks are being positioned at different distances -- possibly in a  superposition -- from the mass, or that the mass is positioned at different distances from the clocks~\cite{zych2018relativity}.)

Consider two agents, $\mathrm a$ and $\mathrm b$, with two initially synchronised clocks, each following a fixed world-line.
A third agent prepares one of two mass configurations, $\mathrm{K_{A\prec B}}$ or $\mathrm{K_{B\prec A}}$, so as to induce time dilation between the clocks of $\mathrm a$ and $\mathrm b$. If configuration $\mathrm{K_{A\prec B}}$ is prepared, event $\mathrm A$ -- defined by the clock of agent $\mathrm a$ showing proper time $t_\mathrm{a}=\tau^*$ -- will be in the past light cone of the event $\mathrm B$ which is defined in an analogous way: by the clock of  agent $\mathrm b$ showing proper time $t_\mathrm{b}=\tau^*$. If configuration $\mathrm{K_{B\prec A}}$ is prepared, event $\mathrm B$ will be in the past light cone of event $\mathrm A$. 
To keep the world lines of the agents independent of the mass configuration, their laboratories can be embedded in tight enough trapping potentials i.e.~much stronger than the gravitational field (which is feasible since our protocol does not require macroscopic source masses, see Methods).
In Supplementary Note 4  we discuss other mass configurations which have the desired effect on temporal order but for which the agents $\mathrm{a, b}$ can remain inertial.

\begin{figure}[h!]
\centering
\includegraphics[width=11cm]{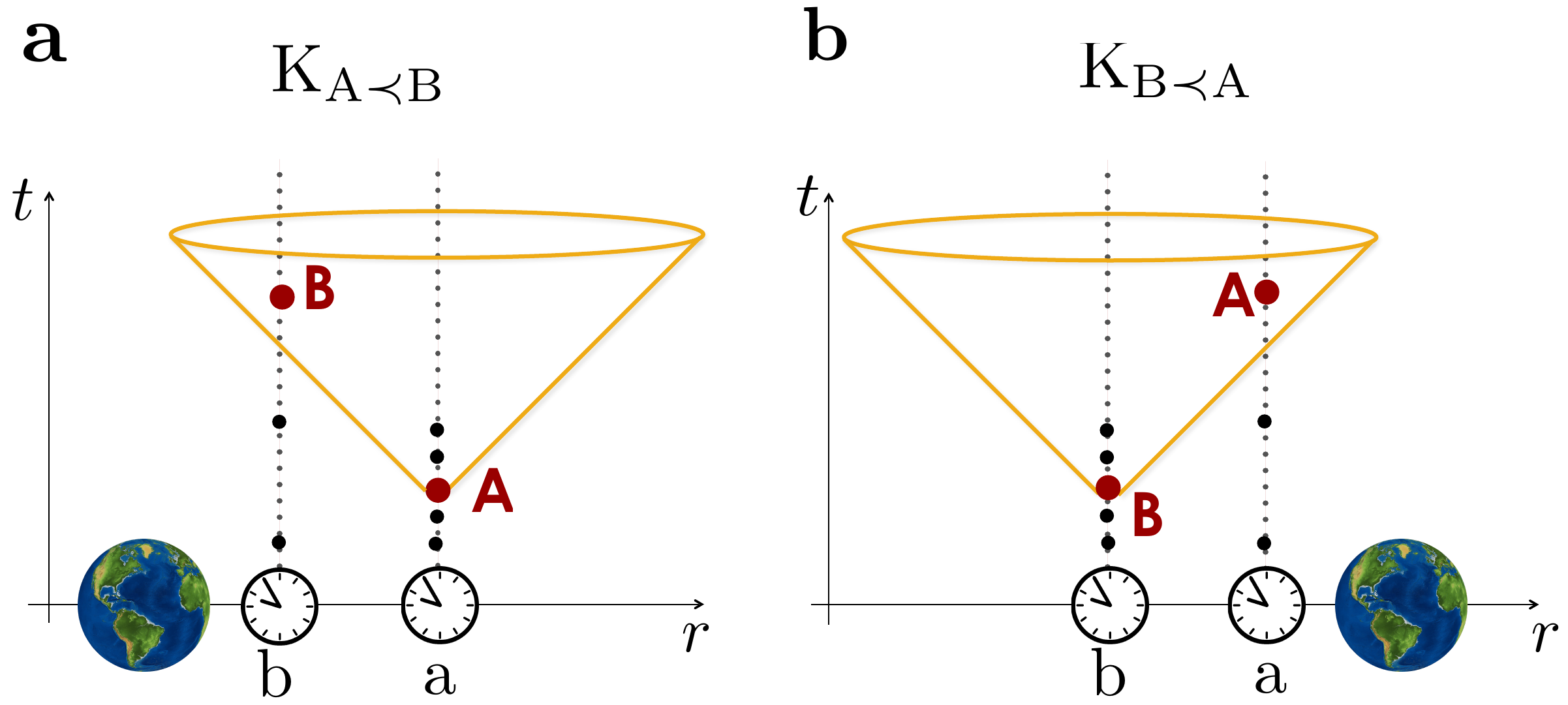}
\caption{General relativistic engineering of causal relations between space-time events using a massive body. Initially synchronised clocks $\mathrm a$ and $\mathrm b$ are positioned at fixed distances from a far-away agent whose time coordinate is $t$. Event $\mathrm{A (B)}$ is defined by the clock of $\mathrm{a (b)}$ showing proper time $\tau^*$. \textbf{a} In configuration $\mathrm{K_{A\prec B}}$ the mass is placed closer to $\mathrm b$ than to $\mathrm a$. Due to gravitational time dilation, event $\mathrm A$ can end up in a causal past of event $\mathrm B$: for a sufficiently large $\tau^*$ the time difference between the clocks becomes greater than it takes light to travel between them. Light emitted at event $ \mathrm A$ reaches clock $\mathrm b$ before the event $\mathrm B$ occurs. \textbf{b} Configuration $\mathrm{K_{B\prec A}}$ is fully analogous to $\mathrm{K_{A\prec B}}$: the mass is placed closer to clock $\mathrm a$ and the event $\mathrm B$ can end up in the causal past of the event $\mathrm A$. \hspace*{\fill}}
\label{event_order}
\end{figure}

A possible way to realise configuration $\mathrm{K_{A\prec B}}$ is to place an approximately point-like body of mass $M$ closer to $\mathrm b$ than to $\mathrm a$, see Fig.\ \ref{event_order}.
The light-cone structure of the resulting space-time is fully determined by the metric tensor $g_{\mu \nu}$, for which we adopt the sign convention (-,+,+,+). In isotropic coordinates in the first-order post-Newtonian expansion the metric components are~\cite{WeinbergGR} $g_\mathrm{00}(r)=-\left(1+2\frac{\Phi(r)}{c^2}\right)$ and $g_{ij}(r)=\delta_{ij}\left(1+2\frac{\Phi(r)}{c^2}\right)^{-1}$, $i,j=1,2,3$, where $\Phi(r)=-\frac{G M}{r}$ is the gravitational potential and $r$ is the spatial distance between the mass and the event where the metric is evaluated. For an event with a spatial coordinate $\mathbf{R}_\mathrm{a}$
and the mass at a spatial coordinate $\mathbf{r}_M$ (where the spatial coordinates are defined e.g.~by a far-away agent as in Fig.~\ref{event_order}) we have $r\equiv| \mathbf{R}_\mathrm{a}-\mathbf{r}_M|$. Note that we use a common coordinate system to describe the different mass configurations and the associated space-time metrics. Operationally, we can associate such coordinates with the far-away agent, whose local clocks are not affected by the change in the matter distribution. However, this is only a convenient interpretation,  we can always think of the coordinates in analogy to gauge fixing -- any physical prediction regarding proper times of the clocks and exchange of the signals will not depend on the choice of coordinates.

We consider that $\mathrm a$ and $\mathrm b$ remain at fixed coordinate distances from the mass, $r_\mathrm{a}$ and $r_\mathrm{b}  = r_\mathrm{a} - h$ respectively, and find the parameters for which event $\mathrm A$ ends up in the past light-cone of $\mathrm B$ for $\mathrm{K_{A\prec B}}$ (and vice versa for  $\mathrm{K_{B\prec A}}$). An infinitesimal proper time element along a world line at a distance $r$ from the mass is given by $d\tau(r)=\sqrt{-\go(r)}dt,$ where $t$ is the coordinate time, and a photon travelling in the radial direction from $r_\mathrm{a}$ reaches $r_\mathrm{b}$ after a coordinate time $T_\mathrm{c}=\frac{1}{c}\int_{r_\mathrm{b}}^{r_\mathrm{a}}dr'\sqrt{-\frac{g_{rr}(r')}{\go(r')}}$. Therefore, if the photon is emitted at the local time $t_\mathrm{a}=\tau^*$, it reaches $r_\mathrm{b}$ when $\mathrm b$'s local time is $\bar t_\mathrm{b}=\sqrt{-g_\mathrm{00}(r_\mathrm{b})}(\frac{\tau^*}{\sqrt{-\go(r_\mathrm{a})}}+T_\mathrm{c})$, assuming that the local clocks are synchronised so that $t_\mathrm{a}=0$ and $t_\mathrm{b}=0$ coincide with the coordinate time $t=0$. For
\be
\tau^*>T_\mathrm{c}\frac{\sqrt{-\go(r_\mathrm{b})}}{1-\sqrt{\frac{\go(r_\mathrm{b})}{\go(r_\mathrm{a})}}}
\ee{cond_tau}
we have $\bar t_\mathrm{b}\leq\tau^*$, which means that there is enough time for a not-faster-than-light signal emitted at event $\mathrm A$ (defined by $t_\mathrm{a}=\tau^*$) to travel the distance $h$ and reach agent $ b$ at event $ B$ (defined by $t_\mathrm{b}=\tau^*$). This means that event $\mathrm A$  is in the causal past of event $\mathrm B$ as required. For example, for $h\ll r_\mathrm{a}$ condition \eqref{cond_tau} is satisfied for  $\tau^*>\frac{2r_\mathrm{a}^2 c}{GM}$.  Configuration $\mathrm{K_\mathrm{B\prec A}}$ can be arranged  analogously, by placing the mass closer to $\mathrm a$ than to $\mathrm b$. Then, the condition $\tau^*>\frac{2r_\mathrm{b}^2 c}{GM}$, for $h\ll r_\mathrm{b}$, ensures that $\mathrm B$ is in the causal past of $\mathrm A$. Note that with the above conditions on $\tau^*$ the events $\mathrm A$ and $\mathrm B$ are always time-like separated but have different time orders for the two mass configurations -- these conditions guarantee that the time order between $\mathrm A$ and $\mathrm B$ is swapped in all reference frames.

The example above simply illustrates that in general relativity causal structure is dynamical and depends on the stress-energy tensor of the matter degrees of freedom: Preparing different matter distributions on a space-like hypersurface can result in different causal relations between events in its causal future.

\subsection*{Quantum control of temporal order}
\label{qcontrol}
When $\mathrm A$ is in the past light-cone of $\mathrm B$, a physical system can in principle be transferred from $\mathrm A$ to $\mathrm B$. Consider a quantum system $\mathrm S$ initially prepared in state $|\psi\ket^\mathrm{S}$ which undergoes a unitary $U_\mathrm{A}$ at event $\mathrm A$ ({at the space-time location where the clock of agent a marks proper time $\tau^*$}) and a unitary $U_\mathrm{B}$ at event $\mathrm B$. Such ordered events can therefore result in the following state of $\mathrm S$:
\begin{equation}
|\widetilde{\psi}_1\ket^\mathrm{S} = U_\mathrm{B}U_\mathrm{A} |\psi\ket^\mathrm{S}.
\label{AB}
\end{equation}
If $\mathrm B$ is before $\mathrm A$, and $\mathrm S$ is prepared in the same initial state, the final state of $\mathrm S$ is
\begin{equation}
|\widetilde{\psi}_2\ket^S = U_\mathrm{A} U_\mathrm{B} |\psi\ket^\mathrm{S}.
\label{BA}
\end{equation}
A situation can therefore be arranged such that state \eqref{AB} is produced for configuration $\mathrm{K_{A\prec B}}$ and \eqref{BA} is produced for $\mathrm{K_{B\prec A}}$. (We ignore a possible additional time evolution between the two events for simplicity.) Different mass configurations can result in different temporal orders of local operations, which holds in quantum as well as in classical theory. 
Let us make the following assumptions:
\begin{enumerate}
\item[a)] Macroscopically distinguishable states of physical systems can be assigned orthogonal quantum states.
\item[b)] Gravitational time dilation in a 
classical limit reduces to that predicted by general relativity.
\item[c)]\label{superpos} The quantum superposition principle holds (regardless of the mass or nature of the involved system).
\end{enumerate}
Even though the above assumptions hold in the standard quantum and general relativistic frameworks,  it is not known if a fundamental theory of quantum gravity satisfies them. Our aim is to investigate their consequences for the notion of temporal order.

The coordinates introduced in the previous section define a foliation of space-time into equal-time slices. As long as no horizons are present in any of the considered configurations, such slices define space-like hypersurfaces.  With each hypersurface one can associate a Hilbert space, containing the quantum states of interest at the given time.  The time coordinate corresponds to the time $t$ in Fig.~\ref{event_order} and is operationally defined as the time measured by the local clock of the far-away agent (not affected by the mass configurations).
These quantum states can be understood operationally as states assigned by the far away agent. However, as discussed in the previous section, such an interpretation is not strictly necessary but is merely a convenient way to define the relevant mathematical objects and to carry out the calculations.

The two mass configurations $\mathrm{K_{A\prec B}}, \mathrm{K_{B\prec A}}$ can thus be assigned quantum states $|\mathrm{K_{A\prec B}}\ket^\mathrm M$, $|\mathrm{K_{B\prec A}}\ket^\mathrm M$. By assumption $a)$ these states are orthogonal. Since each state {individually} satisfies the classical limit (mass 
is sufficiently localised around a single world-line), following assumption $b)$, the system $\mathrm S$ will evolve as in eq.~\eqref{AB} or \eqref{BA} depending whether the mass is in state $|\mathrm{K_{A\prec B}}\ket^\mathrm M$ or $|\mathrm{K_{B\prec A}}\ket^\mathrm M$, respectively. Finally, by assumption $c)$, a superposition $|\mathrm{K_{+}}\ket^\mathrm M := \frac{1}{\sqrt{2}}\left( |\mathrm{K_{A\prec B}}\ket^\mathrm M + |\mathrm{K_{B\prec A}}\ket^\mathrm M\right)$ is a physically {allowed} mass configuration, and will yield the following final state of the joint system:
\begin{equation}
|\psi_{\textrm{sup}}\ket^{\mathrm{MS}} = \frac{1}{\sqrt{2}}\left(|\mathrm{K_{A\prec B}}\ket^\mathrm M U_\mathrm{B}U_\mathrm{A}|\psi\ket^\mathrm{S} + |\mathrm{K_{B\prec A}}\ket^\mathrm M U_\mathrm{A}U_\mathrm{B}|\psi\ket^\mathrm{S}\right).
\label{superposition}
\end{equation}
An explicit calculation showing how this state arises is presented in Methods. 
We note that not only classical gravity but also semi-classical~\cite{kiefer2012quantum} and stochastic gravity~\cite{Hu_Verdaguer:2003} theories would not yield eq.~\eqref{superposition} since these frameworks describe gravitational interactions in terms of classical, possibly stochastic, variables, thus violating assumption c).

{Note that, given a specific physical system used as a clock, it is possible to simulate its time dilation using non-gravitational interactions. For example, an electric field can shift atomic energy levels and thus ``time dilate'' a clock based on atomic transitions. Therefore, one can produce a state analogous to \eqref{superposition} without using gravity. However, only gravity can alter the relative ordering of events independently of the nature of the systems and interactions used as clocks, due to the universality of time dilation: The preparation and manipulation of the massive object can be carried out without any knowledge of other aspects of the protocol. Such a universality underpins a fundamental distinction between our gravitational protocol and other, non-gravitational, methods to control causal relations between operationally-defined events~\cite{procopio_experimental_2014, Rubinoe1602589, rubino2017experimental, Goswami2018, goswami2018communicating, Wei2019, guo2018experimental}. (See also Supplementary Note 4 for further discussion.)} 

{Finally, the} state \eqref{superposition} is the result of a process wherein the order of operations on a target system (S) is determined by the quantum state of a control system (position of the massive body). {Such a process} is known as a {quantum switch} \cite{Chiribella:2013} and has been studied as a possible quantum-information resource \cite{Chiribella:2012, Colnaghi:2012, Araujo:2014, feixquantum2015, guerin16}. The state $|\psi_{\textrm{sup}}\ket^{\mathrm{MS}}$ is a superposition of two amplitudes corresponding to different {predefined, classical} orders between events $\mathrm A$ and $\mathrm B$.
Note that, if the control system is discarded, the reduced state of $\mathrm S$ is
\begin{equation}
\frac{1}{2}\left(|\widetilde{\psi}_1\ket\bra \widetilde{\psi}_1|^\mathrm{S} + |\widetilde{\psi}_2\ket\bra \widetilde{\psi}_2|^\mathrm{S}\right),
\label{separable}
\end{equation}
which is indistinguishable from a {probabilistic mixture} of $|\widetilde{\psi}_1\ket$ and $|\widetilde{\psi}_2\ket$. The state in  Eq.\ \eqref{separable} can be interpreted as arising from events A and B with a classical, albeit unknown, temporal order. Therefore, any protocol aimed at testing operationally quantum features of temporal order necessarily requires a measurement of the control system.

\subsection*{Bell's theorem for temporal order}

The above argument shows that superpositions of massive objects can {in principle} result in a coherent quantum control of temporal order between events. However, one might question whether such a conclusion has a direct physical meaning or whether it relies on a particular interpretation of state \eqref{superposition}. Furthermore, the state assignment is defined in terms of a given coordinate system, while we would like to base our conclusions on coordinate-independent physical events. Since the very meaning of quantum states and measurements might be put into question in the absence of a classical space-time, a proof of non-classical causal relations should not rely on the validity of the quantum formalism.  In the following we show that it is possible to probe the nature of temporal order irrespective of the validity of quantum theory.  We formulate a theory independent argument -- which does not rely on the quantum framework and provides means to exclude the very possibility of explaining data from a hypothetical experiment in terms of a  classical temporal order (which can be stochastic and dynamical) within a broad class of probabilistic theories, not limited to quantum mechanics. 
Our formulation is analogous to  Bell's  theorem for local hidden variables \cite{bell64, Clauser1969} (see Methods) and we thus refer to the theorem below as Bell's theorem for temporal order of events. 
The core of the argument is simple: Given a bipartite system prepared in a separable state, it is not possible to violate any bipartite Bell inequality by performing local operations (transformations and measurements) on the two parts, as long as the local operations are applied in a definite order.

\begin{figure}[h!]
\centering
\includegraphics[width=5.5cm]{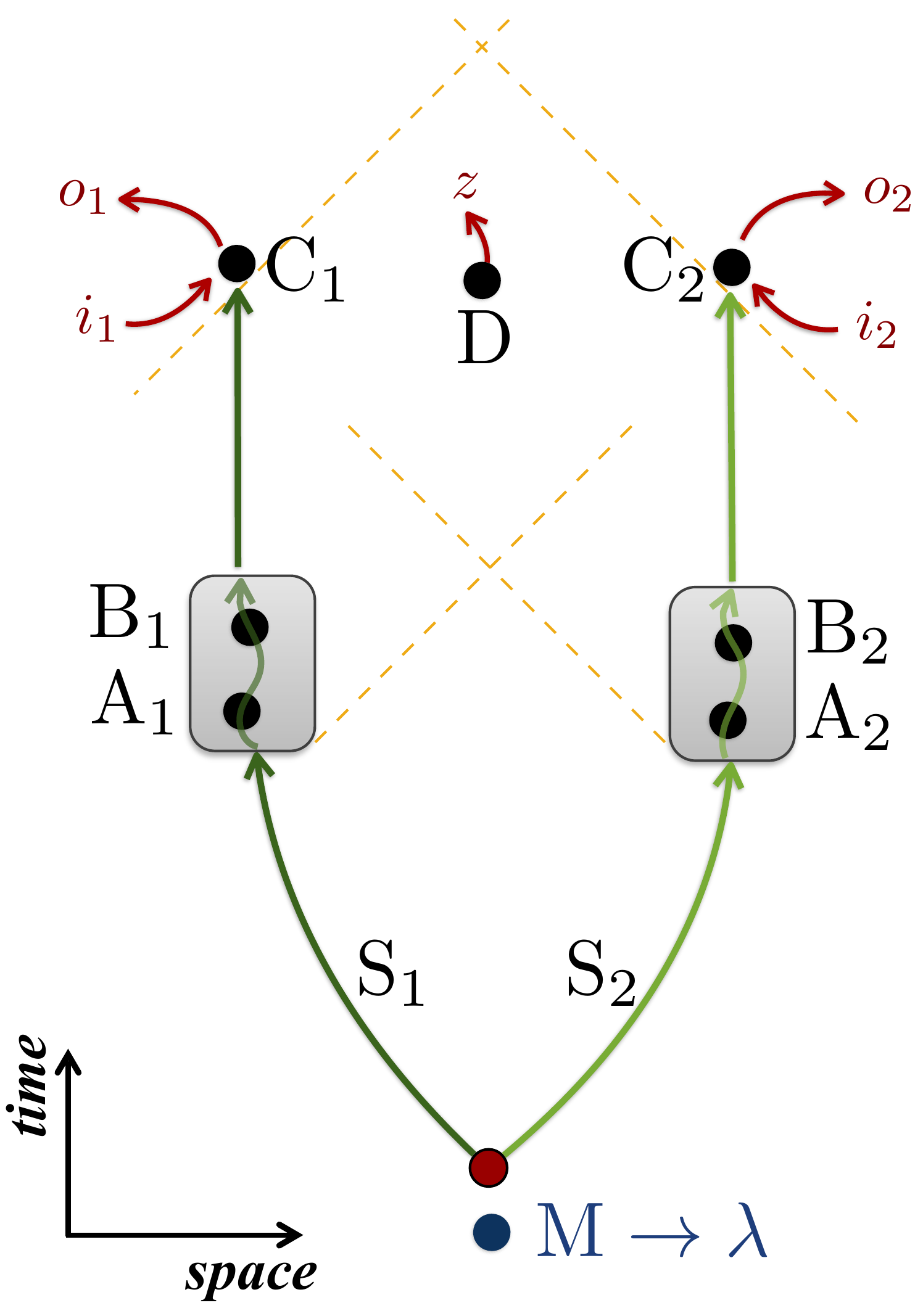}
\caption{Bell's theorem for temporal order. A bipartite system, made of subsystems $\mathrm S_1$ and $\mathrm S_2$, is sent to two groups of agents. Operations on $\mathrm S_1$ ($\mathrm S_2$) are performed at events $\mathrm A_1$, $\mathrm B_1$ ($\mathrm A_2$, $\mathrm B_2$). At event $\mathrm C_1$ ($\mathrm C_2$), a measurement with setting $i_1$ ($i_2$) and outcome $o_1$ ($o_2$) is performed. Events $\mathrm A_1$, $\mathrm B_1$ are space-like separated from $\mathrm A_2$, $\mathrm B_2$ and $\mathrm C_1$ is space-like to $\mathrm C_2$; light cones are marked by dashed yellow lines. The order of events $\mathrm A_j$, $\mathrm B_j$, $j=1,2$, is described by a variable $\lambda$ defined by a system $\mathrm M$. The system $\mathrm M$ is measured at event $\mathrm D$, producing an output bit $z$.
If the initial state of the systems $\mathrm{S_1, S_2, M}$ is separable, and $\lambda$ is a classical variable (possibly dynamical and probabilistic),  the resulting bipartite statistics of the outcomes $o_1, o_2$ cannot violate any Bell inequality, even if conditioned on $z$.\hspace*{\fill}}
\label{bell_for_time_1}
\end{figure}

The scenario 
involves a bipartite system with subsystems $\mathrm S_1$ and $\mathrm S_2$ and a system $\mathrm M$ that can influence the temporal order of events. For  $j=1,2$, each system $\mathrm{S}_j$ undergoes two transformations, $T_{\mathrm{A}_j}$ and $T_{\mathrm{B}_j}$, at space-time events $\mathrm A_j$, $\mathrm B_j$, respectively. Each system is then measured at an event $\mathrm C_j$ according to some measurement setting $i_j$, producing a measurement outcome $o_j$.
Additionally, $\mathrm M$ is measured at an event $\mathrm D$, space-like separated from both $\mathrm C_1$ and $\mathrm C_2$, producing an outcome $z$, see Fig.\ \ref{bell_for_time_1}.
We now define the notion of classical order between events: 

\textit{Definition 1,} {A set of events is {classically ordered} if,
for each pair of events $\mathrm A$ and $\mathrm B$, there exist a space-like surface and a classical variable $\lambda$ defined on it that determines the causal relation between $\mathrm A$ and $\mathrm B$: for each given $\lambda$, either $\mathrm{A \preceq B}$ ($\mathrm A$ in the past causal cone of $\mathrm B$), $\mathrm{B \preceq A}$ ($\mathrm A$ in the past causal cone of $\mathrm B$), or $\mathrm{A || B}$ ($\mathrm A$ and $\mathrm B$ space-like separated).}

Classically ordered events do not necessarily form a partially ordered set: classical order can be dynamical (the order between two events can depend on some operation performed in the past, i.e.\ some agent can prepare $\lambda$) and stochastic ($\lambda$ might be distributed according to some probability, and not specified deterministically)~\cite{Oreshkov2015, abbott2016}. 

\textbf{Bell's theorem for temporal order.} No states, set of transformations
and measurements which obey assumptions  \ref{s}---\ref{free} below can result in a violation of the Bell inequalities.
\begin{enumerate}[1)]
	\item\label{s} {Local state:} The initial state $\omega$ of $\mathrm{S_1, S_2}$ and $\mathrm M$ is separable (as defined in Methods).
	\item\label{ssl} {Local operations:} All transformations performed on the systems are local (as defined in Methods).
	\item\label{to} {Classical order:} The events at which operations (transformations and measurements) are performed are classically ordered.
	\item\label{sep}{Space-like separation:} {Events ($\mathrm{A_1, B_1}$) are space-like separated from events ($\mathrm{A_2, B_2}$); $\mathrm{C_1, C_2}$, and $\mathrm D$ are pair-wise space-like separated.}
	\item\label{free} {Free-choice:} The measurement choices in the Bell measurement are independent of the rest of the experiment (This is a standard assumption necessary in Bell-like theorems).
\end{enumerate}

More formally, let us denote by $\mathbb{T}={(T_\mathrm{A_1}, T_\mathrm{B_1}, T_\mathrm{A_2},  T_\mathrm{B_2})}$ the set of all local transformations irrespective of their order. The thesis of the theorem can be rephrased as: the conditional probability
\begin{equation}
P\left(o_1,o_2|i_1,i_2,z,{\mathbb{T}},\omega\right)
\label{conditional}
\end{equation}
produced under assumptions \ref{s}--\ref{free} does not violate Bell's inequalities for any value of $z$. {The proof of the theorem is presented in Methods.}

%


\subsection*{Violation of Bell inequalities for temporal order}
\label{sec:violation}

\begin{figure}[ht!]
\centering
\includegraphics[width=9.0cm]{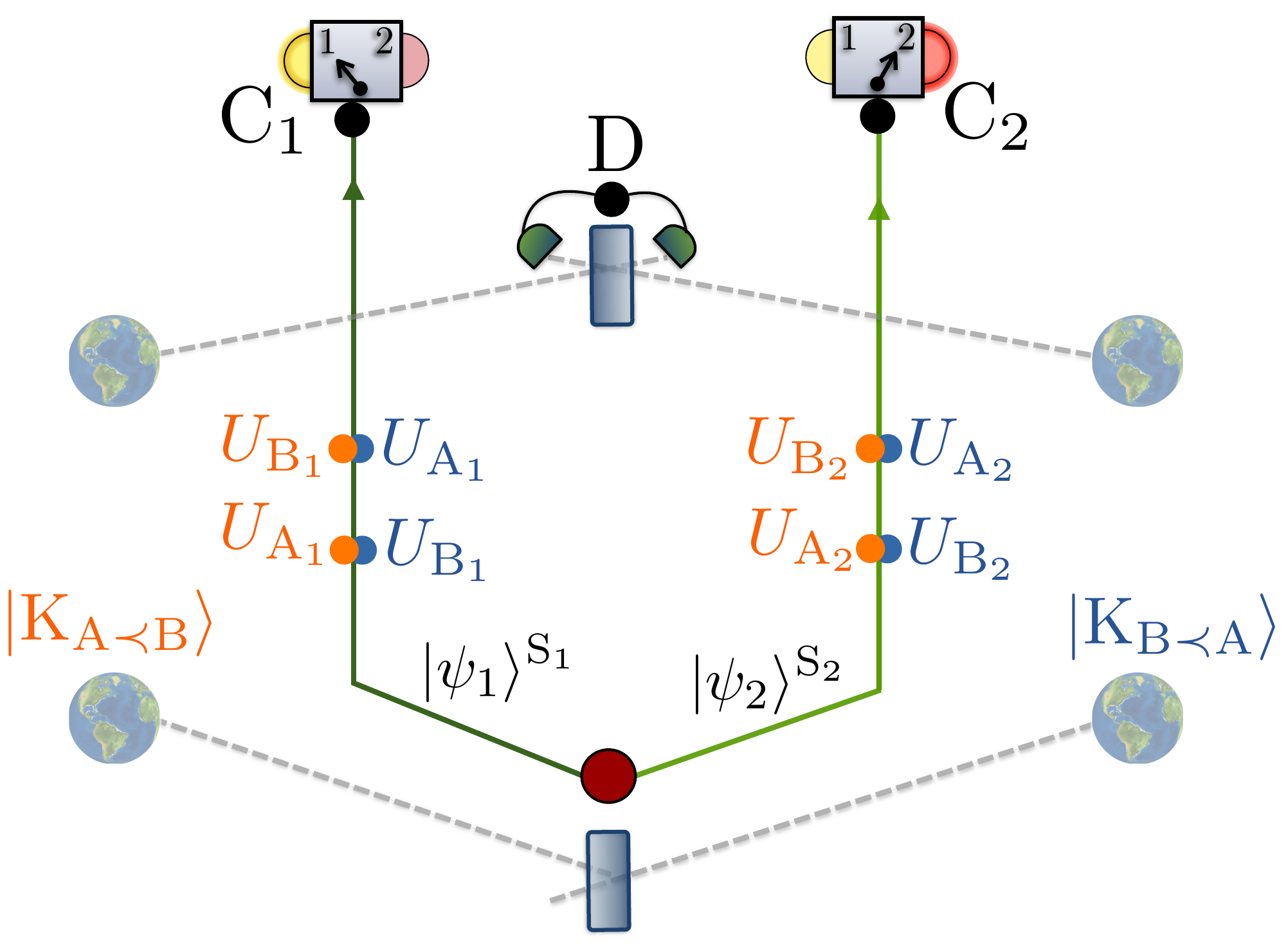}
\caption{Schematics of a protocol for a violation of Bell's inequalities for temporal order. Systems $\mathrm{S_1, S_2}$ are prepared in a product state $|\psi_1\ket^\mathrm{S_1}|\psi_2\ket^\mathrm{S_2}$ and sent to space-like separated regions. One pair of agents performs unitary operations $U_\mathrm{A_1}$, $U_\mathrm{B_1}$ on $\mathrm S_1$ at the correspondingly marked space-time events; another pair acts on $\mathrm S_2$ with unitary operations $U_\mathrm{A_2}$ and $U_\mathrm{B_2}$. Each operation is applied only once, at an event defined by the specific proper time of the local clock of the agent. A massive body is prepared in a superposition of two configurations $|\mathrm{K_{A\prec B}}\ket$ and $|K_{B\prec A}\ket$ which define different causal structures for future events. For the amplitude $|\mathrm{K_{A\prec B}}\ket$, the operations  $U_{\mathrm{A}_i}$ $i=1,2$ applied on $\mathrm S_i$ are in the causal past of the operations $U_{\mathrm{B}_i}$ (orange dots); and vice versa for $|\mathrm{K_{B\prec A}}\ket$ (blue dots). The operations can be chosen such that $U_{\mathrm{A}_i}U_{\mathrm{B}_i}|\psi_i\ket^{\mathrm S_i}$ is orthogonal to $U_{\mathrm{B}_i}U_{\mathrm{A}_i}|\psi_i\ket^{\mathrm{S}_i}$ (for both $i=1$ and $i=2$), resulting in a maximally entangled final state.  Bell measurements are performed at events $\mathrm C_1$ and $\mathrm C_2$ on $\mathrm S_1$ and $\mathrm S_2$, respectively.  At event $\mathrm D$  the mass is measured in a superposition basis. Conditioned on the outcome this measurement,  the results of the measurements at $\mathrm C_1$, $\mathrm C_2$ can maximally violate Bell's inequalities, which would not be possible if the order of events was classical (even if probabilistic).\hspace*{\fill}}
\label{bell_for_time}
\end{figure}

Here we show how the gravitational quantum control of temporal order from Sec.\ \ref{qcontrol} can result in events whose temporal order is entangled: 
A bipartite quantum system, initially in a product state $|\psi_1\ket^\mathrm{S_1}|\psi_2\ket^\mathrm{S_2}$, is sent to two different regions of space such that $\mathrm{a_1}$, $\mathrm{b_1}$, and c$_1$ only interact with S$_1$, while a$_2$, b$_2$, and c$_2$ only interact with S$_2$. Agents a$_1$, a$_2$ perform respectively the unitaries $U_\mathrm{A_1}$, $U_\mathrm{A_2}$ at the events A$_1$, A$_2$, while agents b$_1$, b$_2$, perform the unitaries $U_\mathrm{B_1}$, $U_\mathrm{B_2}$ at the events B$_1$, B$_2$. Finally, $\mathrm{c_1}$ and $\mathrm{c_2}$ measure S$_1$ and S$_2$ at events C$_1$ and C$_2$, respectively, see Fig.\ \ref{bell_for_time}.
Assume that a massive system can be prepared in two configurations, $\mathrm{K_{A\prec B}}$ and $\mathrm{K_{B\prec A}}$, such that $\mathrm{A_1 \prec B_1 \prec C_1}$ ($\mathrm{ A_1}$ in the past light-cone of B$_1$, etc) and $\mathrm{A_2 \prec B_2 \prec C_2}$ for $\mathrm{K_{A\prec B}}$, while $\mathrm{B_1 \prec A_1 \prec C_1}$ and $\mathrm{B_2 \prec A_2 \prec C_2}$ for $\mathrm{K_{B\prec A}}$; and such that the events are space-like separated as per assumption \ref{sep}, which can always be achieved by having the groups sufficiently separated. If the mass is prepared in superposition $\frac{1}{\sqrt{2}}\left(|\mathrm{K_{A\prec B}}\ket^{M} + |\mathrm{K_{B\prec A}}\ket^{M}\right)$, the joint  state of the mass and the systems after the application of the unitaries is
\begin{equation}
\frac{1}{\sqrt{2}}\left(|\mathrm{K_{A\prec B}}\ket^\mathrm{M} U_\mathrm{B_1}U_\mathrm{A_1}|\psi_1\ket^\mathrm{S_1} U_\mathrm{B_2}U_\mathrm{A_2}|\psi_2\ket^\mathrm{S_2} + |\mathrm{K_{B\prec A}}\ket^\mathrm{M} U_\mathrm{A_1}U_\mathrm{B_1}|\psi_1\ket^\mathrm{S_1} U_\mathrm{A_2}U_\mathrm{B_2}|\psi_2\ket^\mathrm{S_2}\right).
\end{equation}
Agent $\mathrm d$ at the event $\mathrm D$ measures the mass in the superposition basis $|\pm\ket = \frac{1}{\sqrt{2}}\left( |\mathrm{K_{A\prec B}}\ket \pm |\mathrm{K_{B\prec A}}\ket\right)$. Conditioned on the outcome, the joint state of $\mathrm S_1$ and $\mathrm S_2$ reads
\begin{equation}
\label{entangled}
\frac{1}{\sqrt{2}}\left(U_\mathrm{B_1}U_\mathrm{A_1}|\psi_1\ket^\mathrm{S_1} U_\mathrm{B_2}U_\mathrm{A_2}|\psi_2\ket^\mathrm{S_2} \pm U_\mathrm{A_1}U_\mathrm{B_1}|\psi_1\ket^\mathrm{S_1} U_\mathrm{A_2}U_\mathrm{B_2}|\psi_2\ket^\mathrm{S_2}\right).
\end{equation}
If the states $U_\mathrm{B_1}U_\mathrm{A_1}|\psi_1\ket^\mathrm{S_1}$, $U_\mathrm{B_2}U_\mathrm{A_2}|\psi_2\ket^\mathrm{S_2}$ are orthogonal to $U_\mathrm{A_1}U_\mathrm{B_1}|\psi_1\ket^\mathrm{S_1}$, $U_\mathrm{A_2}U_\mathrm{B_2}|\psi_2\ket^\mathrm{S_2}$, respectively, then the state \eqref{entangled} is maximally entangled. Local measurements can thus be performed on subsystems $\mathrm S_1$, $\mathrm S_2$ whose outcomes will violate Bell inequalities, conditioned on the measurement outcome at $ \mathrm D$ (see Supplementary Note 2 for an example).

The above {thought experiment} can in principle be realised in a scenario where it is meaningful to argue that assumptions \ref{s}, \ref{ssl} and \ref{sep}, \ref{free} are satisfied.  Violation of the Bell's inequality would then imply that assumption \ref{to} does not hold,  proving non-classicality of temporal order.  In order to maximally violate the inequality,  the time-dilated clocks of the agents need to decorrelate from the systems $\mathrm S_i$. In Methods we present a particular scenario using photons that satisfies also this requirement.
In Supplementary Note 3 we present two concrete examples of our thougth experiment, using  as the systems $\mathrm S_i$ polarisation states of photons, depicted in Supplementary Figure 1, or spatial modes of a quantum field, depicted in Supplementary Figure 2.

\section*{Discussion}\label{sec:discussion}
The non-classical causal structures discussed in this work arise in a semi-classical, albeit non-perturbative, regime where no {explicit} quantization of the gravitational field is needed (which is complementary to the regime of most quantum gravity frameworks \cite{kiefer2012quantum}).
Our approach shows that general relativity and standard quantum mechanics are sufficient to analyse scenarios involving superpositions of macroscopically different {classical} backgrounds. Not only  is there no tension between the two frameworks, but there is also {no ambiguity} in the prediction of physical effects that arise: For each probability amplitude the time-dilation effects introduced by the mass can be treated classically. The considered processes involve a simple superposition of such amplitudes and the final probability amplitude is given by the usual Feynman sum. Note that, even though no explicit quantisation of the metric is used, the amplitudes in the Feynman sum do correspond to macroscopically distinct space time metrics: This is because each of these amplitudes contains a different causal structure, which determines the metric up to a conformal factor~\cite{Hawking1976, Malament1977}. {Quantisation of the metric is therefore implicit in our result, in a similar way as in recently considered witnesses for quantum gravity in interferometric scenarios \cite{bose2017spin, marletto2017entanglement, belenchia2018quantum}.}


A practical realization of the Bell-test for time order would be extremely challenging, even in the light of current efforts to prepare superposition states of massive objects and test their gravitational interactions~\cite{Bose:1999:PRA, marshall2003towards, ref:Kleckner2009, KimBoseUlbricht:2016PRL, Schmoele:2016}. However, assumption that the violation is {fundamentally} impossible has far reaching consequences: this {would imply} that time order, and thus time itself, {can} be described with a classical parameter even in space-times originating from a quantum state of a massive object -- with no need to invoke any other mechanism, such as refs~\cite{ref:Karolyhazy1966, ref:Diosi1989, ref:Penrose1996, Stamp:2012, Penrose2014},   that would decohere these states (see also Supplementary Note 5 for further discussion of such an assumption). On the other hand, since these mechanisms postulate a specific decoherence time of spatial superpositions, one could think that they preclude the preparation of non-classical causal structures. This is not the case: the time required to complete our protocol can be shorter than the decoherence time postulated by these models (see Methods). Thus, contrary to some motivations \cite{ref:Penrose1996, Penrose2014}, these models do not enforce {fundamentally} classical space-time with a fixed causal structure {(i.e.~there is a parameter regime where entangled causal structures could form but decoherence postulated by these models is negligible).}
Finally, classical temporal order could not be excluded 
also in a scenario where massive bodies can be prepared in quantum states but one (or more) of the assumptions \ref{s}, \ref{ssl}, \ref{sep}, \ref{free} cannot be satisfied for some fundamental reason. We note that in particular the notion of locality may be fundamentally limited in the context of quantum gravity~\cite{PhysRevD.74.064018, PhysRevD.93.024030}.

We should note that proof-of-principle realisations of indefinite causal order, analogous to the examples discussed here, have been realised in the laboratory. However, such realisations cannot be interpreted as proofs of non-classical space-time in the sense of general relativity, see Supplementary Note 4 for a discussion of the key differences between the gravitational and other methods for a quantum control of temporal order. The full extent of the relation between gravitational and non-gravitational realisations of quantum causal structures merits an in-depth study that goes beyond the scope of the present work and will be explored elsewhere.

A crucial aspect of Bell's theorem for temporal order is that it provides a theory independent result -- it applies to any framework where causal relations are described classically, such as classical, semi-classical~\cite{kiefer2012quantum} and stochastic gravity~\cite{Hu_Verdaguer:2003} theories. 
Moreover, joint validity of the quantum superposition principle and gravitational time dilation, assumptions a)--c),   suffice for a maximal possible violation of the bound. Therefore, a classical notion of temporal order is untenable in any theory compatible with these basic principles. Finally, the way in which a non-classical causal structure can be engineered exploiting time dilation from a massive body in a quantum state reveals a close connection between the information-theoretic framework of quantum combs/process matrices  and joint effects of quantum mechanics and general relativity. 

\section*{Methods}\label{methods}

\subsection*{Quantum Gravitational control of temporal order.}\label{QG-SWITCH}
According to the Einstein equations, a massive object 
gives rise to a spacetime metric $g_{\mu\nu}$, $\mu, \nu=0,...,3$ which in isotropic coordinates and a post-Newtonian expansion reads~\cite{WeinbergGR}: $g_\mathrm{00}(r)=-\big(1+2\frac{\Phi(r)}{c^2}\big)$,
$g_{ij}(r)=\delta_{ij}\big(1-2\frac{\Phi(r)}{c^2}\big),$ $i,j=1,2,3$
where $r$ denotes the distance to the location of the mass. In other words, if a test mass or a clock is positioned at a spatial coordinate $\mathbf{R}_\mathrm{a}$ as described by a far away agent (as in Fig.~\ref{event_order})
and the massive object is at a coordinate $\mathbf{r}_M$, then $r = | \mathbf{R}_\mathrm{a}-\mathbf{r}_M|$ which for clarity we denote below by $R_\mathrm{a}-{r}_M$. It is important to note that the same metric and coordinates can describe scenarios  where the mass is placed at different locations at a finite distance from $\mathbf{r}_M$, as long as it remains far from an asymptotic region so that the spatial and temporal coordinates of the far away agent remain unaffected (i.e.\ are those of flat Minkowski space time). 
In these coordinates the Hamiltonian of a clock -- a particle with internal degrees of freedom -- reads 
\be
H_\mathrm{a}=\sqrt{-g_\mathrm{00}(R_\mathrm{a}-r_M)\big(\Omega_\mathrm{a}^2+c^2g_{ij}(R_\mathrm{a}-r_M)P^iP^j\big)},
\ee{MethodHam}
(see e.g.~\cite{zych2015quantumEEP, Castro:2017EntangledClocks,zych2018gravitational}) where $P^i$, $i=1,2,3$ are the components of the momentum operator, and $\Omega_\mathrm{a}$ is the internal Hamiltonian, 
describing {the} local time evolution of the internal degrees of freedom. 
Note that we can restrict ourself to an effectively one-dimensional scenario, so only one of the spatial coordinates has been kept in the above expression. In the first post-Newtonian expansion and considering that both the mass and the clock follow fixed world lines at constant $\mathbf R_\mathrm{a}$ and $\mathbf r_M$, respectively, eq.~\eqref{MethodHam} becomes
\be
H_\mathrm{a}\approx \Omega_\mathrm{a}\big(1+\frac{\Phi(R_\mathrm{a}-r_M)}{c^2}\big).
\ee{HA_approx}

The asymptotic time coordinate $t$ defines space-like hypersurfaces that are independent of the location of the mass and on which one can define states of all the involved systems (the clocks, the target systems and the mass itself) and Hamiltonian \eqref{HA_approx} describes their time evolution of with respect to $t$. Due to the interactions between the mass and the clocks  -- effected by the space-time metric which contains the potential $\Phi(R_\mathrm{a}-r_M)$ -- the time evolution of the clocks depends on their relative distance $R_\mathrm{a}-r_M$ to the mass.  Crucially, by the definition of $t$ and the Hamiltonian our description includes both considered different mass configurations: the mass can be semi-classically localised around a single spatial coordinate $r$ or in superposition of different spatial coordinates and the associated states belong to the same Hilbert space associated with a space-like hypersurface labelled by $t$. We thus  have all the  tools to analyse time evolution in the presence of a superposition state of the mass, even though it leads to a quantifiably non-classical causal structure.

With respect to $t$ and the associated foliation of space-time, the evolution of the clock which at $t=0$ is in an internal state $|{s_\mathrm{a}(\tau_0)}\ket$, where $\tau_0$ denotes the clock's proper time at $t=0$, reads 
\be
e^{-i\Omega_\mathrm{a}t\big(1+\frac{\Phi(R_\mathrm{a}-r_M)}{c^2}\big)}|R_\mathrm{a}\ket|{s_\mathrm{a}(\tau_0)}\ket=|R_\mathrm{a}\ket|{s_\mathrm{a}(\tau_0+\tau(R_\mathrm{a}-r_M, t))}\ket,
\ee{MethodTimeEvolution} 
where $\tau(R_\mathrm{a}-r_M, t) := t\big(1+\frac{\Phi(R_\mathrm{a}-r_M)}{c^2}\big)$ is the proper time elapsing for the clock at a radial distance $|R_\mathrm{a}-r_M|$ from the mass when the elapsed coordinate time is $t$;  and for clarity we set $\hbar=1$.

Before continuing on to the gravitational quantum control, we give an example of an internal Hamiltonian, state, and evolution. Let us take  $\Omega_\mathrm{a} =E_0|0\ket\bra0|+E_1|1\ket\bra1|$ and $|s_\mathrm{a}(\tau_0=0)\ket=\frac{1}{\sqrt{2}}(|0\ket+|1\ket)$, which describe e.g.~an atom in an equal superposition of some two electronic energy levels $|0\ket, |1\ket$ with energies $E_0, E_1$, respectively. Under $H_\mathrm{a}$ from eq.~\eqref{HA_approx} internal state $|s_a(0)\ket$ from eq.~\eqref{MethodTimeEvolution} evolves as $$e^{-i\Omega_\mathrm{a}t\big(1+\frac{\Phi(R_\mathrm{a}-r_M)}{c^2}\big)}|s_a(0)\ket=\frac{1}{\sqrt{2}}e^{-iE_0t\big(1+\frac{\Phi(R_\mathrm{a}-r_M)}{c^2}\big)}|0\ket+\frac{1}{\sqrt{2}}e^{-iE_1t\big(1+\frac{\Phi(R_\mathrm{a}-r_M)}{c^2}\big)}|1\ket$$
$$\equiv \frac{1}{\sqrt{2}}e^{-iE_0\tau(R_\mathrm{a}-r_M, t)}|0\ket+\frac{1}{\sqrt{2}}e^{-iE_1\tau(R_\mathrm{a}-r_M, t)}|1\ket,$$ which is simply  $|s_\mathrm{a}(\tau(R_\mathrm{a}-r_M, t))\ket$.

We now use the above to show how the quantum superposition principle and general relativity lead to the prediction that quantised matter acts as a quantum control of temporal order. To this end we assume conditions a)--c) from Sec.~\ref{qcontrol}  and consider two clocks  positioned at $R_\mathrm{A}$ and $R_\mathrm{B}$, respectively. The Hamiltonian of clock a is thus eq.~\eqref{HA_approx} and fully analogously for b, $H_\mathrm{b}\approx \Omega_\mathrm{b}\big(1+\frac{\Phi(R_\mathrm{b}-r_M)}{c^2}\big)$. The clocks are initially synchronised with each other and with a clock of the distant agent so that at $t_0=0$ both clocks are at $\tau_0=0$. We further consider a target system, e.g. a mode of the electromagnetic field, initially in a state $|\psi\ket^\mathrm{S}$, on which an operation $\mathcal{O}_\mathrm{A}$ is performed  at an event $\mathrm A=(R_\mathrm{a}, \tau_\mathrm{a}=\tau^*)$ and an operation $\mathcal{O}_\mathrm{B}$ at an event $\mathrm{B}=(R_\mathrm{b}, \tau_\mathrm{b}=\tau^*)$, where $\tau_\mathrm{a}$, $\tau_\mathrm{b}$ refer to the proper {times} of the clock A, B, respectively. We effectively represent these operations as $\mathcal{O}_\mathrm{A}=\delta(\tau_\mathrm{a}-\tau^*, r-R_\mathrm{a})\mathrm{O_A}$, where $\delta(\tau_\mathrm{A}-\tau^*, r-R_\mathrm{a})$ is a Dirac delta distribution and $\mathrm{O_A}$ is an operator (e.g.~describing rotation of the polarisation of an electromagnetic field mode by a particular half-wave plate) independent of time and location.  The total Hamiltonian reads
 \be
H_\mathrm{tot}=H_\mathrm{a}+H_\mathrm{b}+\mathcal{O}_\mathrm{A}+\mathcal{O}_\mathrm{B},
 \ee{totalHam}
which for simplicity assumes trivial time evolution of the mass and of the target system between the application of the operations. We  furthermore consider the following initial (at $t_0=0$) state of the mass, clocks and the target system:
\be
|\psi(0)\ket^\mathrm{MSab}=|R_\mathrm{a}\ket|R_\mathrm{b}\ket|s_\mathrm{a}(\tau_0=0)\ket|s_\mathrm{b}(\tau_0=0)\ket|\psi\ket^\mathrm{S}\left(|r_\mathrm{L}\ket^\mathrm{M} + |r_\mathrm{R}\ket^\mathrm{M}\right),
\ee{joint_instate}
where positions $r_\mathrm{L}$, $r_\mathrm{R}$ of the mass refer to the configurations in the left and the right panel of Figure~\ref{event_order}, respectively, i.e.~they realise configurations $\mathrm{K_{A\prec B}}$ and $\mathrm{K_{B\prec A}}$: for $|r_\mathrm{L}\ket$ the mass is at a distance $r_\mathrm{a}=r_\mathrm{L}-R_\mathrm{a}$ from clock a and at $r_\mathrm{b}=r_\mathrm{a}-h$ from $\mathrm b$, while for $|r_\mathrm{R}\ket$  the relative distances are swapped and the mass is at a distance $r_\mathrm{a}-h$ 
from $ \mathrm a$ and at $r_\mathrm{a}$ from $\mathrm b$.
After coordinate time $t$ such that $\tau(r_\mathrm{a}, t)>\tau^*$ (where $\tau^*>\frac{2r_\mathrm{b}^2 c}{GM}$, see main text) the state evolves to
\be
\begin{split}
 |\psi(t)\ket^\mathrm{MSab}=|R_\mathrm{a}\ket|R_\mathrm{b}\ket&\bigg(|s_\mathrm{a}(\tau(r_\mathrm{a}, t))\ket|s_\mathrm{b}(\tau(r_\mathrm{a}-h,t))\ket e^{-i\mathrm{O}_B}e^{-i\mathrm{O}_\mathrm{A}}|\psi\ket^\mathrm{S}|r_\mathrm{L}\ket^\mathrm{M} \\
 +&|s_\mathrm{a}(\tau(r_\mathrm{a}-h, t))\ket|s_\mathrm{b}(\tau(r_\mathrm{a},t))\ket e^{-i\mathrm{O_A}}e^{-i\mathrm{O_B}}|\psi\ket^\mathrm{S}|r_\mathrm{R}\ket^\mathrm{M}\bigg).
 \end{split}
 \ee{state_switch1}
 The order of applying unitary  transformations $U_\mathrm{A}=e^{-i\mathrm{O_A}}$ and $U_\mathrm{B}=e^{-i\mathrm{ O_B}}$ to the target system is controlled by the position of the mass, which due to time dilation changes causal relations between  events $ \mathrm A$ and $ \mathrm B$. Swapping the mass distribution: $|r_\mathrm{L}\ket\rightarrow|r_\mathrm{R}\ket$, $|r_\mathrm{R}\ket\rightarrow|r_\mathrm{L}\ket$ and letting the state evolve for another time interval $t$ results in the final state where the clocks become synchronised again
 \be
\begin{split}
 |\psi(t)\ket^\mathrm{MSab}=&|R_\mathrm{a}\ket|R_\mathrm{b}\ket|s_\mathrm{a}(\tau_\mathrm{f})\ket|s_\mathrm{b}(\tau_\mathrm{f})\ket
 \bigg( U_\mathrm{B}U_\mathrm{A}|\psi\ket^\mathrm{S}|r_\mathrm{R}\ket^\mathrm{M} + U_\mathrm{A}U_\mathrm{B}|\psi\ket^\mathrm{S}|r_\mathrm{L}\ket^\mathrm{M}\bigg),
 \end{split}
 \ee{state_switch2}
where $\tau_\mathrm{f}=\tau(r_\mathrm{a}, t)+\tau(r_\mathrm{a}-h, t)$. Measuring the mass in a superposition basis $|r_\mathrm{L}\ket^\mathrm{M} \pm |r_\mathrm{R}\ket^\mathrm{M}$ prepares the target system in the corresponding superposition state  $U_\mathrm{B}U_\mathrm{A}|\psi\ket^\mathrm{S} \pm U_\mathrm{A}U_\mathrm{B}|\psi\ket^\mathrm{S}$.

The above example demonstrates that under very conservative assumptions a spatial superposition of a mass generates a quantum-controlled application of unitary operations. More fundamentally, this effect {stems from the} 
superposition of different causal structures associated with the superposed states of the mass.


\subsection*{Proof of Bell's theorem for temporal order}\label{proof}

{Bell's theorem {in general} asserts that, under certain assumptions,  {the correlations between the} outcomes of independent measurements on two subsystems {must} satisfy a class of inequalities. The two measuring {parties} are referred to as Alice and Bob. {In every experimental run}, each of them measures one of two properties of the subsystem {they receive. For each of the properties, one of two outcomes is obtained, for convenience chosen to be $\pm 1$}.
{Bell's inequalities follow from the the conjunction of the following assumptions: }
1) measurement results are determined by properties that exist prior to and independent of the experiment (hidden variables); 2) results obtained at one location are independent of any measurements or actions performed at space-like separation (locality); 3) any process that leads to the choice {of} which measurement will be carried out is independent from other processes in the experiment (free choice). The outcomes of Alice $A(i, \lambda)$ and Bob $B(i, \lambda)$  thus only depend on their own choice of setting, index $i$,  and on the property of the system, variable $\lambda$. The correlation between outcomes $A(i, \lambda)$ and $B(i, \lambda)$ for the measurement choices $i,j$ is described by $E(A_i,B_j) = \int d\lambda P(\lambda) A(i,\lambda)B(j,\lambda)$, where $P(\lambda)$ is the probability distribution over the properties of the systems.
It is straight-forward to check that {one possible} inequality satisfied by the correlations $E(A_i, B_j)$ is the so-called Clauser-Horne-Shimony-Holt (CHSH) inequality: $|E(A_1,B_1) + E(A_1,B_2) + E(A_2,B_1)-E(A_2,B_2)|\leq 2$. 
{Crucially, quantum theory allows for the left-hand-side of this inequality to reach a value larger than $2$, and experimental measurements of this (and other inequalities) have confirmed such a violation \cite{Clauser:1972, hensen2015experimental, Giustina2015, Shalm2015}.}
The significance of the violations of  Bell's inequalities is in showing that neither nature nor quantum mechanics obey all three assumptions mentioned above. }

The assumption of {classical order} is sufficient to derive {Causal Inequalities}~\cite{Oreshkov:2012, branciard16}: tasks that, without any further assumptions, cannot be performed on a classical causal structure. However, it is not possible to violate causal inequalities using quantum control of order~\cite{Oreshkov2015, araujo15}, this is why   additional assumptions were required in the present context. It is an open question whether a gravitational implementation of a scenario that does allow for a violation of causal inequalities is possible.

The theorem we have formulated {is} theory independent, but {not} fully device-independent, as it requires the notions of a physical state and a physical transformation (in addition to the measured probability distributions), which we introduce below and then proceed to the proof.
Discussion of the present work in the  context of the theory-dependent framework of causally non-separable quantum processes~\cite{Oreshkov:2012, araujo15, Oreshkov2015} and the fully theory- and device-independent approach of causal inequalities \cite{Oreshkov:2012, branciard16} is presented in Supplementary Note 1. 

We consider a sufficiently broad framework to describe physical systems that can undergo transformations and measurements, similar to generalised probabilistic theories~\cite{hardy2001quantum, Barrett2007, Chiribella2010}. {This framework is more general than quantum or classical theory and we thus need to define  key notions required in the proof.} In this framework, a state $\omega$ is a complete specification of the probabilities $P(o|i,\omega)$ for observing outcome $o$ given that a measurement with setting $i$ is performed on the system.
We are interested in situations where a system can be split up in subsystems, say $\mathrm S_1$ and $\mathrm S_2$, with space-like separated agents performing independent operations on $\mathrm S_1$ and $\mathrm S_2$. We say $\omega$ is a product state, and write $\omega =\omega_1\otimes \omega_2$, if probabilities for local measurements factorise as $P(o_1,o_2|i_1,i_2,\omega)= P(o_1|i_1,\omega_1)P(o_2|i_2,\omega_2)$. If state $\omega_1^f$ is prepared for system $\mathrm S_1$ and state $\omega_2^f$ is prepared for system $\mathrm S_2$, according to a probability distribution $P(f)$ for some variable $f$, we write $\omega=\int\! df\,P(f)\omega_1^f\otimes \omega_2^f$ and say the state is separable. Probabilities  are then given by the corresponding mixture: $P(o_1,o_2|i_1,i_2,\omega)= \int df P(o_1|i_1,\omega_1^f)P(o_2|i_2,\omega_2^f) P(f)$. Note that for such a decomposition Bell inequalities cannot be violated~\cite{bell64, fine82}.

A physical transformation of the system is represented by a function $\omega\mapsto T(\omega)$.
To make our arguments precise we need a notion of local transformations,  namely, realised at the time and location defined by a local clock. If $\mathrm S_1$ is the subsystem on which a local transformation $T_1$ acts, and $\mathrm S_2$ labels the degrees of freedom space-like separated from $T_1$, then, by definition, $T_1$ transforms product states as $\omega_1\otimes\omega_2\mapsto T_1(\omega_1)\otimes \omega_2$ and separable states by convex extension. How local operations act on general, non-separable states can depend on the particular physical theory; however, action on separable states will suffice for our purposes. We further need to define how the local transformations combine. This depends on their relative spatio-temporal locations: If  transformations $T_1, T_2$ are space-like separated they combine as $(T_1\otimes T_2)(\omega_1\otimes \omega_2) = T_1(\omega_1)\otimes T_2(\omega_2)$, which follows from the definition above; if $T_1$ is in the future of $T_2$, we define their combination as $T_1\circ T_2 (\omega) = T_1\left(T_2 (\omega) \right)$. (For simplicity, we omit possible additional transformations taking place between the specified events, as they are of no consequence for our argument).

\begin{proof}
Assumption \eqref{s} says that there is a random variable $f$ determining the local states $\omega_1^f$, $\omega_2^f$ of systems $\mathrm S_1$, $\mathrm S_2$, respectively. Assumption \eqref{to} says there is a random variable $\lambda$ that determines the order of events. In general, the two variables can be correlated by some joint probability distribution $P(\lambda,f)$. By assumption \eqref{sep}, events labelled $A_1, B_1$ are space-like separated from events $A_2, B_2$ and the order between events within each set $(A_j, B_j)$, $j=1,2$ can be defined by a permutation $\sigma_j$. Most generally, there is a probability $P(\sigma_j|\lambda)$ that the permutation $\sigma_j$ is realised for a given $\lambda$.  By assumption \eqref{ssl}, for each given order the system undergoes a transformation {$T^{\sigma_1}\otimes T^{\sigma_2}$}, where {$T^{\sigma_1}$} is the transformation obtained by composing {$T_\mathrm{A_1}$} and {$T_\mathrm{B_1}$} in the order corresponding  to the permutation $\sigma_1$ and similarly for {$T^{\sigma_2}$}. (For example, if $\sigma_1$ corresponds to the order $\mathrm {A_1\prec B_1}$, then {$T^{\sigma_1}=T_\mathrm{B_1}\circ T_\mathrm{A_1}$}.)
Furthermore, at event $\mathrm D$ an outcome $z$ is obtained with a probability {$P(z|\lambda, f, \sigma_1, \sigma_2)$}. Finally, using assumption \eqref{s}, we write the probabilities for all outcomes as
\begin{multline}
\label{post_to}
 P\left(o_1,o_2, z|i_1,i_2, {\mathbb{T}},\omega\right) = \\
\sum_{\sigma_1\sigma_2} \int \!d\lambda \,df P(o_1|i_1,{T^{\sigma_1}}(\omega_1^f))P(o_2|i_2, {T^{\sigma_2}}(\omega^f_2))P(\sigma_1|\lambda) P(\sigma_2|\lambda)P(z|\lambda,f, {\sigma_1, \sigma_2})P(\lambda, f),
\end{multline}
 A simple Bayesian inversion  $P(\sigma_1|\lambda) P(\sigma_2|\lambda)P(z|\lambda,f, \sigma_1, \sigma_2)P(\lambda, f) = P(\lambda, f, \sigma_1, \sigma_2|z)P(z)$, where we used $P(\sigma_j|\lambda)=P(\sigma_j|\lambda,f)$, gives the desired probabilities
\begin{multline}
\label{result}
P\left(o_1,o_2|i_1,i_2, z,  {\mathbb{T}},\omega\right) = \\
 \sum_{\sigma_1\sigma_2} \int \!d\lambda \,df P(o_1|i_1,{T^{\sigma_1}}(\omega_1^f))P(o_2|i_2, {T^{\sigma_2}}(\omega_2^f))P(\lambda, f, \sigma_1, \sigma_2|z) =\\
 \int \!d\tilde{f} P(o_1|i_1,{T^{\sigma_1}})P(o_2|i_2, {T^{\sigma_2}})P(\tilde{f}|z),
\end{multline}
where $\tilde{f}$ is a short-hand for the variables $\lambda, f, \sigma_1, \sigma_2$. The above probability distribution satisfies the hypothesis of Bell's theorem and thus cannot violate any Bell inequality.
\end{proof}

\subsection*{Exemplary scenario realising Bell test for temporal order of events}
\label{app:disentangle}

The protocol allowing for the violation of Bell's inequalities for temporal order exploits correlations between the clocks of the agents $\mathrm{a_1, b_1}$ and the agents $\mathrm{a_2, b_2}$, created due to time dilation induced by the mass. {It should be noted that} the protocol allows maximal violation of the Bell inequality if the joint state of the systems $\mathrm S_1$ and $\mathrm S_2$ is pure {(and maximally entangled)} when the Bell measurements are realised. Thus, for a maximal violation,  the clocks need to decorrelate from the mass  after the application of the unitaries. Below we sketch a scenario that can achieve this.

\begin{figure}[h!]
\includegraphics[width=14cm]{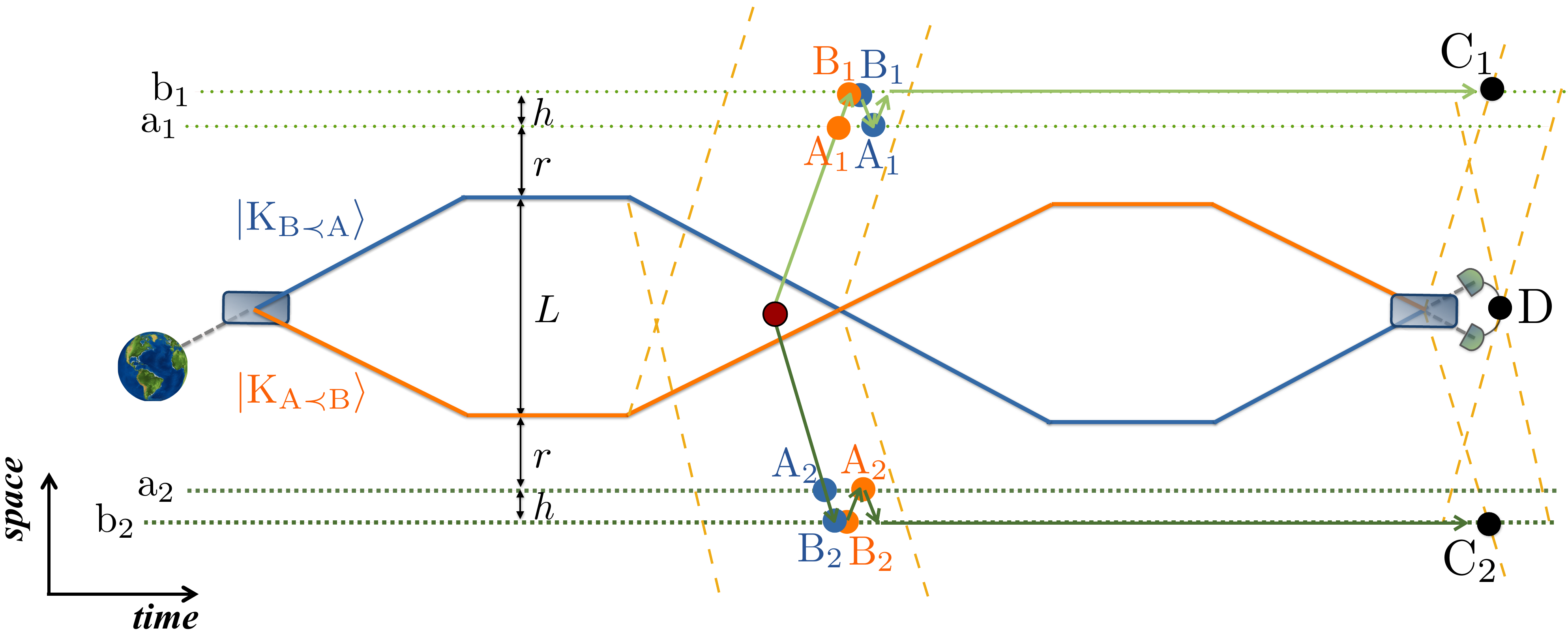}
\caption{Space-time diagram of a protocol for disentangling the clocks from the mass.
In configuration $\mathrm{K_{A\prec B}}$ the mass is at a distance  $r+L$ from $\mathrm a_1$,
and at $r+L+h$ from $\mathrm b_1$. In $\mathrm{K_{B\prec A}}$ -- it is at $r$ from $\mathrm a_1$ and at $r+h$ from $\mathrm b_1$.
The opposite holds for $\mathrm{a_2, b_2}$. The initial mass superposition is swapped (after sufficient time to prepare the clocks in the correlated state) so that they finally show the same time. At the local time $\tau_\mathrm{a}$ of $\mathrm a_1$ (at event $\mathrm A_1$) the agent applies $U_\mathrm{A_1}$  on $\mathrm S_1$. At the local time $\tau_\mathrm{b}$ of $\mathrm b_1$ the agent applies $U_\mathrm{B_1}$ on $\mathrm S_1$.  For the mass configuration  $\mathrm{K_{A\prec B}}$  $\mathrm A_1$ is before $\mathrm B_1$ (orange-coloured events), while for  $\mathrm{K_{B\prec A}}$ event $\mathrm B_1$ is before $\mathrm A_1$ (blue-colored events). The opposite order holds for events $\mathrm{A_2, B_2}$ occurring on the opposite side of the mass, where agents $\mathrm{a_2, b_2}$ act on $\mathrm S_2$. Unitary operations should be applied {in the future light-cone of the event where the clocks get correlated and outside the future light cone of the event when the mass amplitudes are swapped}, Bell measurements (at $\mathrm{C_1, C_2}$) should be made when the clocks become disentangled {(at future light-like events to when the mass amplitudes are brought together)}, and the measurement at event $\mathrm D$ should be space-like to $\mathrm{C_1, C_2}$;  dashed yellow lines represent the relevant light-cones.\hspace*{\fill}}\label{disentangle}
\end{figure}

The space-time arrangement of the mass and the agents in this example is presented in Fig.~\ref{disentangle}. It can be realised in one spatial dimension:  agents acting on the system $\mathrm S_1$ are located at distance $h$ from each other, and the mass is placed at distance $r$ (configuration $\mathrm{K_{B\prec A}}$) or $r+L$ (configuration $\mathrm{K_{A\prec B}}$) from agent $\mathrm a_1$. Agents acting on system $\mathrm S_2$ are placed symmetrically on the opposite side of the mass, such that the mass is at a distance $r+L$ from $\mathrm a_2$ in configuration $\mathrm{K_{B\prec A}}$ and $r$ in configuration $\mathrm{K_{A\prec B}}$. Here, events $\mathrm{A}_j$ are defined by the local time $\tau_\mathrm{a}$ that differs from the local time $\tau_\mathrm{b}$ defining $\mathrm{B}_j$, $j=1,2$. In such a case, even though the mass is always closer to  $\mathrm{a}_j$ than to $\mathrm{b}_j$, the two mass configurations can lead to different event orders -- as they induce different relative time dilations. (Equivalently, one can introduce an initial offset in the synchronisation of the clocks.) Note that the time orders between the two groups  are here ``anti-correlated'': $\mathrm {A_1 \prec B_1}$ and $\mathrm{B_2 \prec A_2}$ for $\mathrm{K_{A\prec B}}$, and vice versa for $\mathrm{K_{B\prec A}}$.
Since otherwise the scenario is the same for $\mathrm S_1$ and $\mathrm S_2$, we focus on the operations performed on $\mathrm S_1$.
The key observation is that swapping the mass distribution, as depicted in Fig.~\ref{disentangle}, will eventually disentangle the clocks from the mass, and since the clocks must be suitably time-dilated when the operations are performed, the operations must not take place in the future light cone of the swapped mass state.

The proper time $\tau_\mathrm{a}$ that has to elapse for the clock of $\mathrm a_1$  such that the order of events  is $\mathrm{A_1 \prec B_1}$ for  $|\mathrm{K_{A\prec B}}\ket$ and  $\mathrm{B_1 \prec A_1}$ for  $|\mathrm{K_{B\prec A}\ket}$ for the present case reads 
\be
\tau_\mathrm{a}= \sqrt{-g_\mathrm{00}(r)}\frac{T_\mathrm{c}(r,h) + T_\mathrm{c}(r+L,h)\sqrt{\frac{g_\mathrm{00}(r+L+h)}{g_\mathrm{00}(r+h)}}}{  1 - \sqrt{\frac{g_\mathrm{00}(r)g_\mathrm{00}(r+L+h)}{g_\mathrm{00}(r+h)g_\mathrm{00}(r+L)}}},
\ee{Method_taua}
where $T_\mathrm{c}(r, L/2)$~is the coordinate travel time of light between radial  distances $r$ and $r+L/2$ from the mass.  The coordinate time corresponding to $\tau_\mathrm{a}$ is $T_\mathrm{a}=\tau_a/\sqrt{-g_\mathrm{00}(r)}$. 
The proper time of event $\mathrm B_1$ is then defined as: 
\be
\tau_\mathrm{b}=\sqrt{-g_\mathrm{00}(r+L+h)}(\frac{\tau_\mathrm{a}}{\sqrt{-g_\mathrm{00}(r+L)}}+ T_\mathrm{c}(r+L,h)).
\ee{Method_taub}
It can directly be checked that when the mass is placed in configuration $\mathrm{K_{A\prec B}}$ -- at a distance $r+L$ from $\mathrm a_1$-- the event $\mathrm A_1$ defined by local clock of $\mathrm a_1$ showing proper time $\tau_\mathrm{a}$ from eq.~\eqref{Method_taua} is in the past light cone of event $\mathrm B_1$, which is defined by the local clock of $\mathrm b_1$ showing proper time $\tau_\mathrm{b}$ from eq.~\eqref{Method_taub}. When the mass is placed in configuration $\mathrm{K_{B\prec A}}$, event $\mathrm B_1$ ends up in the past of the event $\mathrm A_1$.
The coordinate time required for the application of the operations can be estimated as twice the travel time of light between the agents, $T_\mathrm{o}=2T_\mathrm{c}(r+L/2, h)$.

The world lines of the mass can be arranged such that: a) the mass is moving slow so that the two amplitudes of the mass are swapped in a time interval longer than $T_\mathrm{o}$; b) during the application of the operations the distance of each agent to the mass is approximately the same for both mass configurations (as in Fig.~\ref{disentangle}). The first guarantees that  there is enough time to apply the operations after the clocks get correlated, the second -- that the slow-down of light in curved space time, the Shapiro delay \cite{ref:Shapiro1964, ref:Shapiro1971},  can be neglected.

The coordinate-time duration of the entire protocol can be estimated as $T_\mathrm{p}=2T_\mathrm{a}+4L/2c$, where $L/2c$ is the minimal time required to put the mass in superposition of amplitudes separated by the distance $L/2$.  Taking as an example
$M\sim0.1\, \mathrm\mu$g, $L=h\sim0.1\,\mathrm \mu$m, $r\sim1$ fm the protocol in  Fig.~\ref{disentangle} takes $T_\mathrm{p}\sim10$ hours. 
Furthermore, we note that a quantum treatment of the local clocks is central to our protocol since the application of the operations on the target systems is conditioned on the states of the clocks. The time-energy uncertainty \cite{fleming, mandelstam} thus poses a limitation to a single-shot precision with which space-time events can be defined with physical clocks. The optimal clock-state in this context -- evolving the fastest --  is a balanced superposition of energy eigenstates; for an energy gap $\hbar\cdot2\pi\nu_\mathrm{c}$, where $\nu_\mathrm{c}$ is the clock frequency, the smallest time that can be resolved by a single quantum system is the so-called orthogonalisation time~\cite{margolus, kosinski, zielinski} $t_{\perp}=1/2\nu_\mathrm{c}$. For the values of parameters quoted above, the coordinate time difference between the superposed locations of the events $\mathrm A_i$, $i=1,2$ is $\sim10^{-15}$ s, and we thus need a system with frequency $\nu_\mathrm{c}\geq10^{15}$ Hz such as a clock based on optical transitions in ytterbium~\cite{Pizzocaro_2017} or mercury~\cite{Hoyt:PRLabsolute}, which both  give $t_\perp\sim 10^{-16}$ s. While this ideal limit is not reached with practical systems, the resolution of current atomic clocks based on such atoms far exceeds this theoretical bound due to averaging over many atoms, with $2.5\cdot10^{-19}$ uncertainty of the clock frequency recently demonstrated in ref.~\cite{PhysRevLett.120.103201}. We further note that by using $n$ entangled atoms, the orthogonalisation time of the entire system becomes $t_\perp/n$ and can thus be even a few orders of magnitude smaller~\cite{PhysRevLett.117.060506} than required.  Finally,  such atoms have masses $\!\sim\!10^{-25}$ kg and their back-action on the metric produced by $M\!\sim\!10^{-7}$ kg  would thus be negligible. Since the mass difference between the atom in the two involved energy levels is $2\pi\hbar\nu_\mathrm{c}/c^2\sim10^{-35}$ kg also  quantum effects from the clocks' mutual gravitational interactions~\cite{Castro:2017EntangledClocks} can  be neglected.

We conclude that it is in principle possible to achieve the required entanglement of orders, swap the mass distribution so as to finally disentangle the clocks from the mass, and satisfy the locality conditions on the events. Although {a direct  experiment in such a regime is not practical, } the above example surprisingly shows that the  regime  where entangled temporal order arises  is in no way related to the Planck-scale. It is usually assumed that the Planck-scale marks the regime where quantum gravity effects become relevant (first discussed in this context by Bronstein \cite{gorelik1992first}), but this is not the case for the superposition of temporal order.
{In terms of a potential experiment,} one could also take a different (theory-dependent) approach and explore possible witnesses of entangled temporal order~\cite{araujo15}, in analogy to witnesses of entanglement in quantum information theory \cite{TERHAL2000319}. A witness would probe the quantum nature of temporal order indirectly and under further assumptions, but in a relaxed parameter range. Such an approach may lead to more feasible experiments, which will be explored in a future study.

A  spatial superposition state of a mass such as used in our protocol is postulated to decohere in various gravity-inspired collapse models 
\cite{ref:Karolyhazy1966, ref:Diosi1989, ref:Penrose1996, Stamp:2012, Penrose2014} (which thus violate assumption c) in Sec.~\ref{qcontrol}). 
However, even if endorsed, these models do not {immediately} preclude realisation of our protocol: the decoherence time scale in those models is the Diosi-Penrose time \cite{ref:Diosi1989, ref:Penrose1996} $T_\mathrm{DP}=\frac{2 \delta^3 \hbar}{G (ML)^2}$, where $\delta$ is a free parameter. For every value of $\delta$ one can find the mass and the relevant distances ($M, r, L, h$) so that the duration of our entire protocol is shorter than $T_\mathrm{DP}$. For example, following the recent ref.~\cite{Bahrami:2014_PRA} and taking $\delta=10^{-7}$ m, for
$r=10^{10} R_\mathrm{Sch}$, $L=5 r$, $h=r$ and $M=1$~g where $R_\mathrm{Sch}\approx 10^{-30}$~m, the protocol from  Figure \ref{disentangle}  takes $T_\mathrm{p}\approx7\times10^{-18}$~s, while $T_\mathrm{DP}\approx0.5$~s. Taking instead the originally proposed value $\delta=10^{-15}$~m \cite{ref:Diosi1989}, the desired regime is achieved e.g.~for $M=10^{-7}$~kg, $r=10^7 R_\mathrm{Sch}$, $L=5\times10^5r$, $h=10^5r$; with $T_\mathrm{p}\sim10^{-23}$~s and $T_\mathrm{DP}\sim10^{-13}$~s.
Thus the above models in principle still  allow for events with entangled temporal order, and do not enforce the classicality of the causal structure of space-time.


\subsection*{Acknowledgments}
We thank G.~Chiribella,  G.~Milburn, H.~Wiseman, and M.~Vojinovic for feedback. M.Z. and F.C.\ acknowledge support through the Australian Research Council (ARC) Centre of Excellence for  Engineered Quantum Systems (CE 110001013), Discovery Early Career Researcher Awards DE180101443, DE170100712, and the Templeton World Charity Foundation (TWCF 0064/AB38). I.P.\ acknowledges support of the NSF through a grant to ITAMP and the Branco Weiss Fellowship -- Society in Science, administered by the ETH Z\"{u}rich. {\v C}.B.\ acknowledges the support of the Austrian Science Fund (FWF) through the Doctoral Programme CoQuS, the project I-2526-N27 and I-2906, the research platform TURIS, and the \"{O}AW Innovationsfond ``Quantum Regime of Gravitational Source Masses''. This publication  was made possible through the support of a grant from the John Templeton Foundation and from the Foundational Questions Institute (FQXi) Fund. The opinions expressed in this publication are those of the authors and do not necessarily reflect the views of the John Templeton Foundation. F.C.~and M.Z.~acknowledge the traditional owners of the land on which the University of Queensland is situated, the Turrbal and Jagera people.\\

\subsubsection*{Supplementary Note 1: Causally non-separable quantum processes}

Non-classical causal relations can be studied within a recent framework for quantum mechanics with no pre-defined causal structure introduced in ref.~\cite{Oreshkov:2012}.
The starting point of the framework is the notion of {local events} that take place in {local regions}, {with spatial and temporal boundaries of the region defined by local clocks}.
An event is identified with an operation performed in the local region (for example a unitary transformation, or a projection on a given state obtained as the result of a measurement).
A physical scenario, comprising the space-time geometry in which the local regions are embedded, the initial state, and the dynamics connecting the regions, is compactly represented by a {process}---a specification of the probabilities for any possible event/local operation to take place in each region.

At a formal level, a local region $\mathrm X$ is defined by an input Hilbert space $\mathcal{H}^\mathrm{X_I}$ and an output Hilbert space $\mathcal{H}^\mathrm{X_O}$, identified with the quantum degrees of freedom on space-like surfaces on the past and future of $\mathrm X$, respectively. Quantum operations are represented as operators $M^\mathrm{X_I X_O} \in \mathcal{L}(\mathcal{H}^\mathrm{X_I})\otimes \mathcal{L}(\mathcal{H}^\mathrm{X_O})$, where $\mathcal{L}(\mathcal{H})$ is the space of linear operators on the Hilbert space $\mathcal{H}$. Probabilities for events in regions $\mathrm A$, $\mathrm B$,\dots are then given by a generalisation of the Born rule:
\begin{equation}
P(M^\mathrm{A_I A_O}, M^\mathrm{B_I B_O},\dots ) = \tr\left[\left(M^\mathrm{A_I A_O}\otimes M^\mathrm{B_I B_O}\otimes\dots\right)\cdot W^\mathrm{A_I A_O B_I B_O\dots}\right],
\label{processrule}
\end{equation}
where $W^\mathrm{A_I A_O B_I B_O\dots}\in \mathcal{L}(\mathcal{H}^\mathrm{A_I})\otimes \mathcal{L}(\mathcal{H}^\mathrm{A_O})\otimes\mathcal{L}(\mathcal{H}^\mathrm{B_I})\otimes \mathcal{L}(\mathcal{H}^\mathrm{B_O})$ is the {process matrix}.

In this formalism, causal relations between local regions are encoded in the process matrix. For example, the process matrix
\begin{align}
W^\mathrm{A_IA_OB_IB_O}=&\rho^\mathrm{A_I}\otimes [[\id]]^\mathrm{A_O B_I}\otimes \id^\mathrm{B_O}, \textrm{ where}\\
[[\id]]^\mathrm{A_O B_I} :=&  \Ket{\id}\Bra{\id}^\mathrm{A_O B_I} \textrm{ and} \\
\Ket{\id}^\mathrm{A_O B_I}:=& \sum_j |j\ket^\mathrm{A_O} |j\ket^\mathrm{B_I},
\end{align}
represents a situation where an agent at $\mathrm A$ receives a state $\rho$, while the output of $\mathrm A$'s operation is sent to $\mathrm B$ through the identity channel. Such a process is only compatible with the order of events $\mathrm{A\preceq B}$; more general processes compatible with an order of events given by a permutation $\sigma$ are denoted $W^{\sigma}$. If the order is determined by a classical variable $\lambda$, defined in some region in the past of all events, the process matrix has the form
\begin{equation}
W=\int d\lambda W^{\sigma_{\lambda}}P(\lambda)
\label{caussep}
\end{equation}
for some probability distribution $P(\lambda)$.

The question of whether a certain quantum scenario can be embedded in a classical space-time, with a classical order of events, thus reduces to the question whether the corresponding process matrix can be decomposed in a mixture of the form \eqref{caussep}, which we call causally separable. (A more general definition, where the order of future events can depend on past events, is not necessary for our analysis.) The quantum switch, described in the main text, section ``Quantum control of temporal order'', is represented by the process matrix
\begin{align} \label{switch}
W&=|\omega\ket\bra \omega | \\
|\omega\ket &= \frac{1}{\sqrt{2}}\left(|\mathrm{K_{A\prec B}}\ket^\mathrm{M_I}|\mathrm{ABC} \ket + |\mathrm{K_{B\prec A}}\ket^\mathrm{M_I}|\mathrm{BAC} \ket\right), \\ \label{ABC}
|\mathrm{ABC} \ket &= |\psi\ket^\mathrm{A_I}\Ket{\id}^\mathrm{A_OB_I}\Ket{\id}^\mathrm{B_OC_I}, \quad
|\mathrm{BAC} \ket = |\psi\ket^\mathrm{B_I}\Ket{\id}^\mathrm{B_OA_I}\Ket{\id}^\mathrm{A_OC_I},
\end{align}
where $\mathrm{M}$ labels the control system and $\mathrm C$ is the region where the system is measured after the operations in regions $\mathrm A$, $\mathrm B$ are performed.
As shown in ref.~\cite{araujo15}, it is possible to find an experimental procedure, namely a set of operations and measurements for $\mathrm A$, $\mathrm B$, $\mathrm C$, $\mathrm M$, that allows proving the causal non-separability of the switch. However, such a causal witness is both device and theory dependent, namely it relies on the quantum description of the operations performed. Causal inequalities~\cite{Oreshkov:2012, branciard16}, on the other hand, provide a device and theory independent test for causal order; however, no quantum-control of causal order can violate causal inequalities, as proven in refs.~\cite{araujo15, Oreshkov2015},
 and it is an open question whether any physically realisable process can.

The process matrix corresponding to the scenario with entangled temporal orders, introduced in the main text is
\begin{align} \label{entang}
W&=|\varpi\ket\bra \varpi | \\
|\varpi\ket &= \frac{1}{\sqrt{2}}\left(|\mathrm{K_{A\prec B}}\ket^\mathrm{M_I}|\mathrm{A_1B_1C_1} \ket|\mathrm{A_2B_2C_2} \ket + |\mathrm{K_{B\prec A}}\ket^\mathrm{M_I}|\mathrm{B_1A_1C_1} \ket |\mathrm{B_2A_2C_2} \ket\right),
\end{align}
using definitions similar to \eqref{ABC}. Just as for the switch, it is easy to prove that process \eqref{entang} is not causally separable: As it is a rank-one projector, it cannot be decomposed as a non-trivial mixture of orders; Yet it does not describe a process with a definite order, because the signalling relations between parties do not define a partial order. However,   a process of this type cannot be used to violate causal inequalities, see e.g.~ref.~\cite{araujo15}.
The procedure described in the main text can nonetheless prove the causal non-separability of process \eqref{entang} in a theory-independent, albeit device-dependent, way.

\subsubsection*{Supplementary Note 2: State and measurements for the CHSH inequality violation}

Consider a two-qubit system in an initial state $|\psi_1\ket^\mathrm{S_1}\otimes|\psi_2\ket^\mathrm{S_2}\equiv |\mathrm{z+}\ket^\mathrm{S_1}\otimes|\mathrm{z+}\ket^\mathrm{S_2}$ and unitaries %
\begin{equation}
\label{unitaries} U_\mathrm{A_1}=U_\mathrm{A_2}\equiv U_\mathrm{A} =\frac{\id + i\sigma_\mathrm{x}}{\sqrt{2}},\qquad  U_\mathrm{B_1}=U_\mathrm{B_2}\equiv U_\mathrm{B}= \sigma_\mathrm{z},
\end{equation}
with $\sigma_\mathrm{x}$ and $\sigma_\mathrm{z}$ the Pauli matrices. We find $U_\mathrm{A}U_\mathrm{B} = \frac{\sigma_\mathrm{z} + \sigma_\mathrm{y}}{\sqrt{2}}$ and $U_\mathrm{B}U_\mathrm{A} = \frac{\sigma_\mathrm{z} - \sigma_\mathrm{y}}{\sqrt{2}}$, and the final state (eq.~(8) in the main text) reads
\begin{equation}
\label{entangled2}
\frac{1}{\sqrt{2}}\left(|\mathrm{x+}\ket^\mathrm{S_1} |\mathrm{x+}\ket^\mathrm{S_2} \pm |\mathrm{x-}\ket^\mathrm{S_1}|\mathrm{x-}\ket^\mathrm{S_2}\right).
\end{equation}
Importantly,  the sign above depends on the result $|\pm\ket$ of the measurement on the massive system. In order to violate Bell inequalities, agent $\mathrm{c_1}$ measures ${\cal C}_1^0=\frac{\sigma_\mathrm{y}-\sigma_\mathrm{z}}{\sqrt{2}}$ for setting $i_1=0$ and observable ${\cal C}_1^1=\frac{\sigma_\mathrm{y}+\sigma_\mathrm{z}}{\sqrt{2}}$ for $i_1=1$, while agent $\mathrm c_2$ measures observable ${\cal C}_2^0=\sigma_\mathrm{y}$ for setting $i_2=0$ and ${\cal C}_2^1=\sigma_\mathrm{z}$ for $i_2=1$. The expectation value of the CHSH correlation \cite{Clauser1969} with the above measurement choices is
\begin{equation}
\left\langle \mathrm{CHSH} \right\rangle_{\pm} =
\left\langle {\cal C}_1^0\otimes {\cal C}_2^0 + {\cal C}_1^0\otimes {\cal C}_2^1 + {\cal C}_1^1\otimes {\cal C}_2^0 - {\cal C}_1^1\otimes {\cal C}_2^1 \right\rangle_{\pm} = \mp 2 \sqrt{2},
\label{CHSH}
\end{equation}
 for the two outcomes $z=\pm 1$ of the measurement at $\mathrm D$.
This means that conditioned on the value of $z$, the measurements at $\mathrm C_1$ and $\mathrm C_2$ violate the CHSH inequality $\left|\left\langle \mathrm{CHSH} \right\rangle\right| \leq 2$. Importantly,  the settings at $\mathrm C_1$ and $\mathrm C_2$ are independent of $z$ and therefore the three measurements can be performed at space-like separation. The violation of CHSH inequality can be verified once all the data are compared.

\subsubsection*{Supplementary Note 3: Realisations of the protocol}

To achieve large time dilation between a pair of clocks one can use a heavy object or, alternatively, any mass $M$ dense enough to put one of the clocks close to its Schwarzschild radius $R_\mathrm{Sch}:=\frac{2GM}{c^2}$. 
The ticking rate of a clock at $R_\mathrm{Sch}+\epsilon$ with $\epsilon\ll R_\mathrm{Sch}$ differs from the ticking rate of an identical clock at $R_\mathrm{Sch}+l$, $l>\epsilon$, by a factor approximately $
\sqrt{\frac{R_\mathrm{Sch}}{\epsilon}\big(1+2\frac{\Phi(R_\mathrm{Sch}+l)}{c^2}\big)}$, which becomes arbitrarily large for a small $\epsilon$. 
{Practical} realisation of the protocol in Figure 3 
in the main text will nevertheless pose a formidable challenge, but it is in principle possible. Below we give an example. 

\begin{figure}[h!]
\begin{center}
\includegraphics[width=12cm]{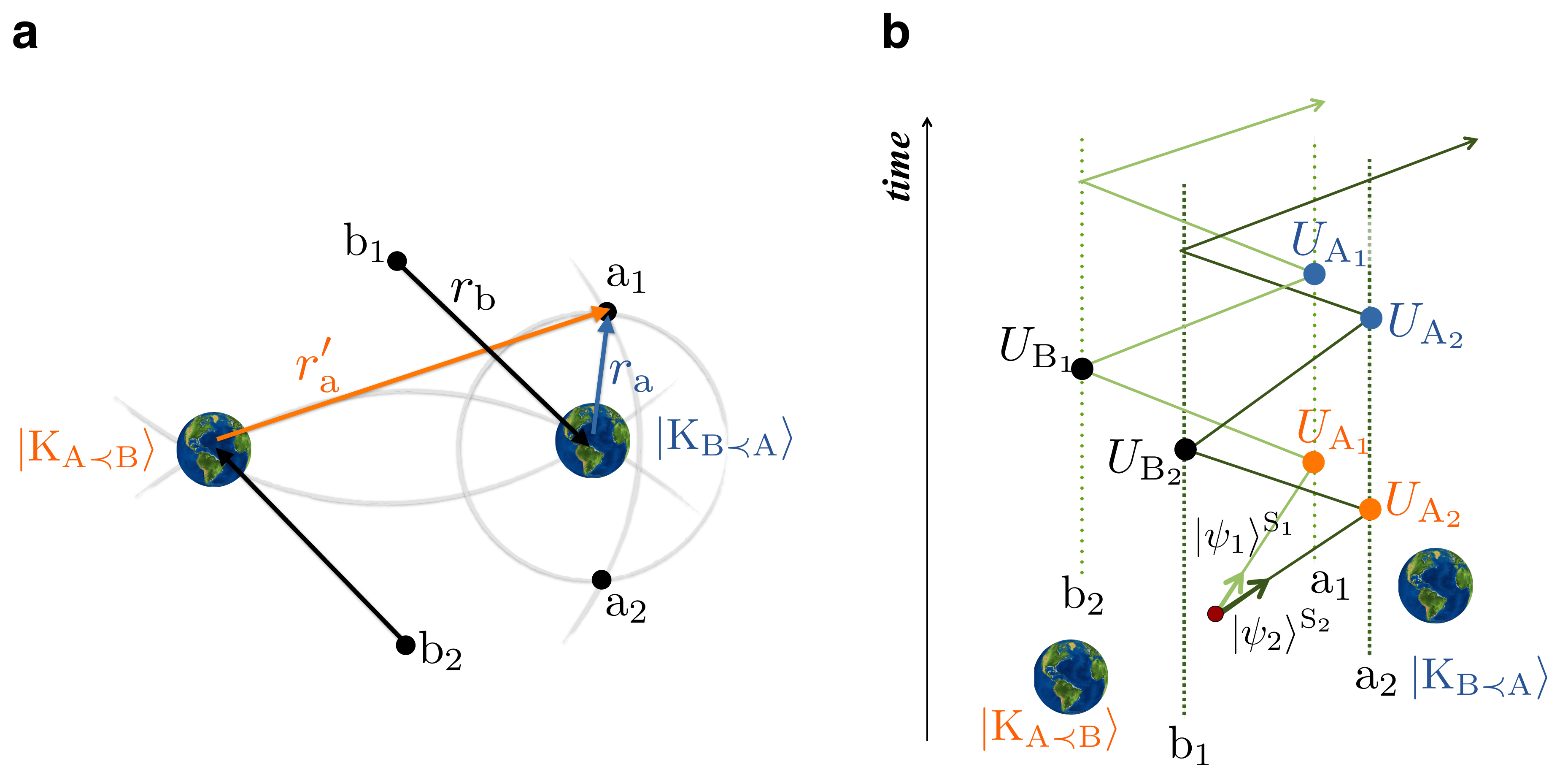}
\caption{Protocol for a violation of Bell's inequalities for temporal order using polarisation states of photons. \textbf{a} Mass configurations $\mathrm{K_{A\prec B}}$, $\mathrm{K_{B\prec A}}$ and location of the agents $\mathrm a_i, \mathrm b_i$, $i=1,2$. $\mathrm b_i$ are at a distance $r_\mathrm{b}$ from both 
{configurations}, while $\mathrm a_i$ are at a distance $r_\mathrm{a}$  from $\mathrm{K_{B\prec A}}$ and $r_\mathrm{a}^\prime>r_\mathrm{a}$ from $\mathrm{K_{A\prec B}}$. \textbf{b} Space-time diagram of the protocol. Systems $\mathrm{S_1, S_2}$ are implemented in the polarisation of two photons, initially in a product state $|\psi_1\ket^\mathrm{S_1}|\psi_2\ket^\mathrm{S_2}$. Green lines are the photons' world lines;  green dotted lines are world lines of the agents. Orange (blue) dots mark events when agents $\mathrm a_i$ apply unitaries $U_{\mathrm A_i}$ for the configuration $\mathrm{K_{A\prec B}}$ ($\mathrm{K_{B\prec A}}$); black dots mark events when $\mathrm b_i$ apply $U_{\mathrm B_i}$. {The photons bounce twice between the agents, but each operation is applied on the photon only once -- when the local clocks of the agents show proper time $\tau^*$. Due to time dilation induced by the mass, $U_{\mathrm A_i}$ are applied before $U_{\mathrm B_i}$ for configuration $\mathrm{K_{A\prec B}}$ (and the events $\mathrm A_i$ coincide with the photons reaching $\mathrm a_i$ for the first time) -- and are applied after $U_{\mathrm B_i}$ for configuration $\mathrm{K_{B\prec A}}$ (and the events $\mathrm A_i$ coincide with the photons reaching $\mathrm a_i$ for the second time). Reprinted
by permission from Springer Customer Service Centre GmbH: Springer International Publishing ``Quantum
Systems under Gravitational Time Dilation'' by M.~Zych (2017).} \hspace*{\fill}}
\label{implementationsBell}
\end{center}
\end{figure}

Consider  configurations $\mathrm{K_{A\prec B}}$, $\mathrm{K_{B\prec A}}$ realised using an effectively point-like body with a fixed mass.
The distance between agents $\mathrm b_i$, $i=1,2$ and the mass is the same for both $\mathrm{K_{A\prec B}}$ and $\mathrm{K_{B\prec A}}$,  while agents $\mathrm a_i$ are closer to the mass in configuration $\mathrm{K_{B\prec A}}$ than in  $\mathrm{K_{A\prec B}}$, see~Fig.~\ref{implementationsBell}~a).
The subsystems  $\mathrm S_i$ can be realised as two identically prepared, uncorrelated photons and the local operations can be performed on their polarisation degrees of freedom (DOF). The photon source is equally distant from $\mathrm{K_{A\prec B}}$ and $\mathrm{K_{B\prec A}}$, (equidistant to the extent that the local clock of  the source remains sufficiently uncorrelated with the mass). All clocks involved in the protocol are initially synchronised with the clock of the source.

At a pre-defined time $T_\mathrm{s}$ according to the clock at the source, the source emits the photon pair -- the emission time is thus uncorrelated with the mass configuration. Photon $\mathrm S_i$ is directed towards agent $\mathrm a_i$, then to $\mathrm b_i$, again to $\mathrm a_i$, back to $\mathrm b_i$,  and exits towards agent $\mathrm c_i$ (or $\mathrm c_i$ simply replaces $\mathrm a_i$), see Fig.\ \ref{implementationsBell} b).
{The agents interact with the relevant DOF of the photon only once, at the time $\tau^*$ as measured by their local clocks.  The unitary transformations are assumed to be independent of the mass configuration (or other aspects of the experiment). As discussed in the main text, the emission time  $T_\mathrm{s}$ of the photons  can be chosen such that event $\mathrm A_i$ (at which  $U_{\mathrm A_i}$ is applied) is before the event $\mathrm B_i$ (at which $U_{\mathrm B_i}$ is applied) and so that event $\mathrm A_i$ coincides with the photon reaching $\mathrm a_i$ for the first time for configuration $\mathrm{K_{A\prec B}}$, and when the photon reaches $\mathrm a_i$ for the second time for $\mathrm{K_{B\prec A}}$. The photon reaches $\mathrm a_i$  twice, before or after $\tau^*$ -- depending on the mass configuration, at which no operation is performed: The photon is reflected with no transformation on the polarisation. The event when the operation $U_{\mathrm B_i}$ is applied always coincides with the photon reaching $\mathrm b_i$ for the first time, since agents $\mathrm b_i$ are at the same distances to the mass for both configurations.}

In general, the travel time of the photon  can depend on the mass configuration due to the Shapiro delay~\cite{ref:Shapiro1964, ref:Shapiro1971}. 
In order to {mitigate} this effect, after the emission time $T_\mathrm{s}$ -- sufficient  to induce the required time dilation between the clocks  -- the mass can be (coherently) moved such that it is at the same distance from each agent (for both $\mathrm{K_{A\prec B}}$ and $\mathrm{K_{B\prec A}}$), or such that it is sufficiently far away from both.
Moreover, in order to de-correlate the time-dilated clocks from the systems $\mathrm S_i$, the amplitudes of the mass can be swapped and the mass can be measured by the agent $\mathrm d$ after a time interval equal to $T_\mathrm{s}$ -- when the clocks of $\mathrm a_i$ and  $\mathrm b_i$ become synchronised again. (We note that Methods section provides details of a protocol that provides both: suppression of the Shapiro effect and decorrelation of the clocks).

Here we discuss another possibility for the realisation of the protocol.
\begin{figure}[h!]
\begin{center}
\includegraphics[width=13cm]{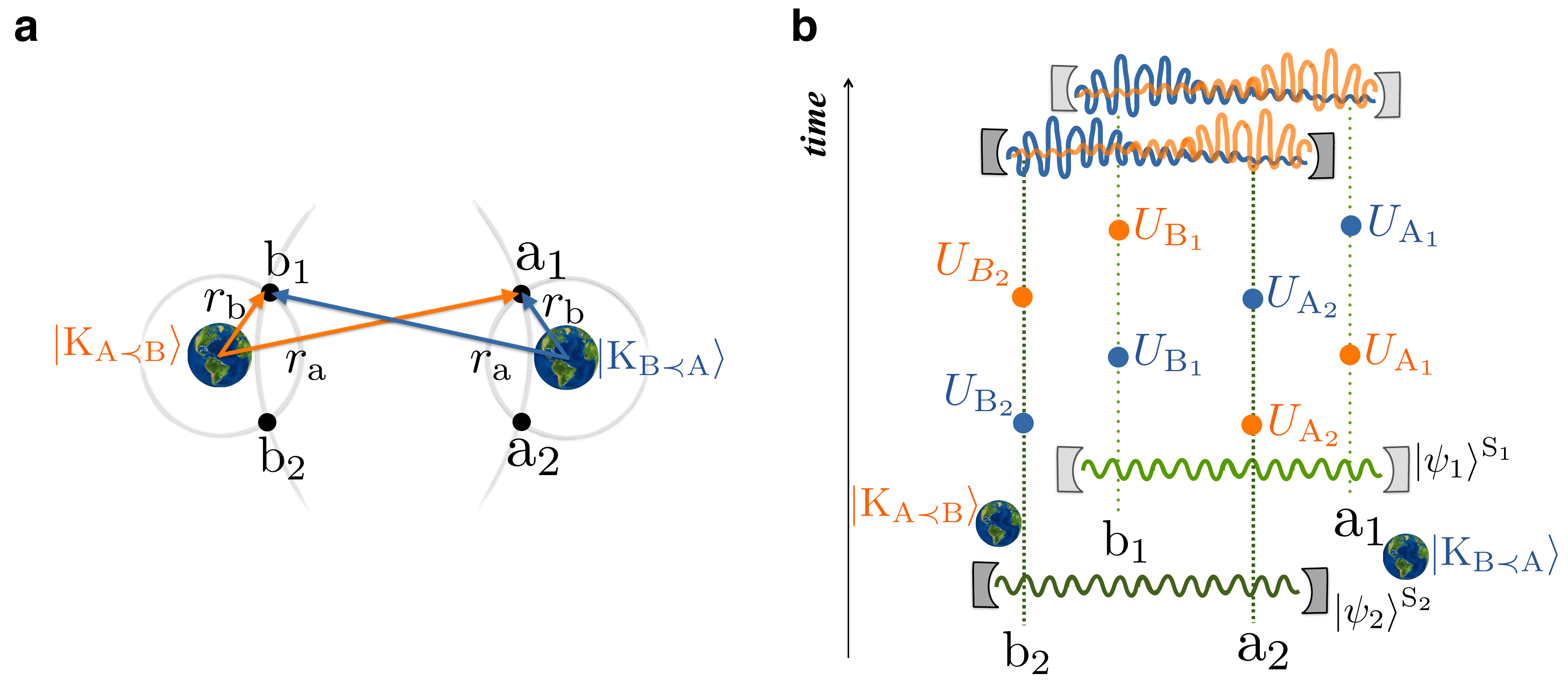}
\caption{Protocol for the violation of Bell's inequalities for temporal order with spatial modes of a quantum field. \textbf{a} Mass configurations: In configuration $\mathrm{K_{A\prec B}}$ the mass is at the same distance $r_\mathrm{b}$ from both $\mathrm b_i$ and at a distance $r_\mathrm{a}>r_\mathrm{b}$ from both $\mathrm a_i$. For the configuration $\mathrm{K_{B\prec A}}$ the mass is at $r_\mathrm{a}$ from both $\mathrm b_i$, and at $r_\mathrm{b}$ from both $\mathrm a_i$. \textbf{b} Space-time diagram. Systems $\mathrm S_i$ are implemented in two regions of an electromagnetic field, prepared in a  vacuum state. Agents apply coarse-grained operations $U_{\mathrm A_i}, U_{\mathrm B_i}$ (in the subspace spanned by the vacuum and a single-photon state) locally on the field at correspondingly marked events. For configuration $\mathrm{K_{A\prec B}}$ ($\mathrm{K_{B\prec A}}$) only orange (blue) events occur, and the final state of the field is represented in orange (blue). Reprinted by permission from Springer Customer Service Centre GmbH: Springer International Publishing ``Quantum Systems under Gravitational Time Dilation'' by M.~Zych (2017).\hspace*{\fill}}
\label{realisation_supp}
\end{center}
\end{figure}
The mass distribution is such that for $\mathrm{K_{A\prec B}}$ the mass is closer to $\mathrm b_1$ than to $\mathrm a_1$ and for $\mathrm{K_{B\prec A}}$ the relative distances are reversed. The same holds for agents $\mathrm{a_2, b_2}$ who are placed symmetrically to agents $\mathrm{a_1, b_1}$ with respect to the mass; see Supplementary Figure~\ref{realisation_supp} a).

The local operations can be performed in the Fock space of a  photon field, more precisely in the two-level subspace spanned by the vacuum and a single-photon state of a chosen field mode. The field is prepared in some mode $\alpha$ at event $\mathrm A_1$, and in mode $\beta$  at  event $\mathrm B_1$. The modes are chosen such,  that the two final states of the field at event $\mathrm C_1$ -- obtained depending on the order between events $\mathrm{A_1, B_1}$  -- are distinguishable. The situation for the agents $\mathrm{a_2, b_2}$ is the same and they prepare the modes $\alpha, \beta$, at the events  $\mathrm{A_2, B_2}$, respectively, see Supplementary Figure~\ref{realisation_supp} b).

One needs to note, that the vacuum state of a relativistic quantum field is entangled with respect to the local subsystems~\cite{ReehSchlieders:1961, SummersWerner:1987part1, SummersWerner:1987part2}. However,  this entanglement is effectively inaccessible under coarse-grained operations~\cite{Zych:2010}. Thus, if the operations performed by the agents are sufficiently coarse-grained, the initial state consisting of the local regions of a vacuum of a quantum field is effectively separable and does not violate the assumptions of the protocol. This implementation 
{differs from the example given in the main text} in that {it does not need a source that would produce a state at a specific time, and distribute it to the agents.}

\subsubsection*{Supplementary Note 4: Gravitation vs other methods for a quantum control of temporal order}

The superposition or entanglement of temporal orders was discussed here in the context 
where a far away agent prepares a quantum state of a massive system which due to relativistic gravity effects yields a desired quantum causal structure for future events. 
There are, however, other methods to control the temporal order between operations applied on a system. One possibility is to control the positions of clocks that define when the operations are applied. For example, by placing clock $\mathrm a$ closer and $\mathrm b$ further away from a fixed mass in superposition with  $\mathrm b$ closer and $\mathrm a$ further, proper times of the clocks become entangled as in the gravitational switch. Quantum control of the time order can also be achieved without any use of gravitational interaction, e.g.~in the extended model of a quantum circuit: Quantum gates can be applied on a system in different orders in a superposition~\cite{Chiribella:2013, Chiribella:2012, Colnaghi:2012, Araujo:2014}
The latter has already been practically implemented using an interferometer to route a photon through two gates (acting on its polarisation) in different orders~\cite{procopio_experimental_2014, Rubinoe1602589, rubino2017experimental, Goswami2018, goswami2018communicating, Wei2019, guo2018experimental}

The key difference between the gravitational scenario presented in this work and other schemes is that, in the latter, the events would be embedded in a classical space-time: In the example of an entangled clock-pair, only these specific clocks could be used to label events for which temporal order is non-classical, while any other nearby clock would define classically ordered events. In the example of an extended quantum circuit, only the photon that went through a beam splitter will undergo different transformations in a non-classical order. In contrast, in the scenario considered in this work \textit{any} local system in the spacetime region affected by the superposition state of the mass will have classically undefined proper time. Thus, any two pairs of clocks in this region will define events with an entangled order.

The above can be highlighted by considering the scenario leading to the violation of the Bell-like inequality for temporal order in a quantum coordinate system \cite{zych2018relativity} defined relative to the massive body, which is here in superposition. (A complete theory of such quantum coordinate transformations is missing; however,  they were also studied within the approach of quantum reference frames in ref.~\cite{Giacomini:2017qcovariance}, where non-relativistic transformations between quantum reference bodies in relative superposition states were constructed such that the dynamics of the systems of interest is described relative to other physical systems, rather than relative to an idealised notion of coordinates.)
 Space-time coordinates of the events would then be defined with respect to the position of the mass and a local clock at its location --  
rather than with respect to the positions and proper times of the clocks of the agents.  By definition, in these coordinates the location of the mass would be fixed. As a result, all local operations performed in the local regions of the agents would appear {to be embedded in a fixed space-time metric} but performed at {different} space-time {events} in superposition -- such that the orders of events in different space-time regions are always entangled. 

Another manifestation of the classical space-time underlying other implementations is that local measurements can reveal the order of events. For the two examples discussed above, this could be achieved, respectively, by non-demolition measurements of
the positions of the clocks or the photon's time of arrival to a gate in a circuit. On the contrary, in the gravitational scenario introduced in this work, all operations are performed at fixed local times, independently of the event order. No local time measurement can reveal whether a given agent is acting first or second. It is further possible to consider a realisation of our scheme where {no local measurement}, temporal or otherwise, can reveal which of the two mass configurations was prepared, and thus what the event order was. This can be achieved by using mass distributions that do not exert any force on the laboratories, but still cause the necessary time dilation.  For example, each laboratory can be placed inside a spherical mass shell~\cite{Hohensee2012}. The gravitational potential inside the shell is constant and depends on the mass and radius of the shell. Therefore, laboratories inside shells of different radii experience relative time dilation but no force if the two shells are sufficiently far from each other.  All steps of the protocol described in the main text can be reproduced using spherical shells: configurations leading to different temporal order are obtained by placing a shell with smaller radius either around laboratory $\mathrm{a}$ or $\mathrm{b}$. The swap of the mass positions would then be replaced by exchange of the shell's radii. This realisation has the further advantage that the laboratories are in free fall and no mass-configuration-dependent acceleration is needed to keep them on the chosen world lines. Note that, even if the gravitational force outside a shells is taken into account, and thus the laboratories might need to accelerate to maintain the desired trajectories, these accelerations do not need to depend on the mass configuration. This is because outside a shell of a fixed mass the same force is exerted at a given distance from its centre-of-mass, independently of the shell's radius. The laboratories would  experience the same gravitational acceleration for each mass configuration, and thus their world lines can still be defined independently of the configuration of mass. For interferometry with massive shells see refs \cite{gooding2015bootstrapping, gooding2014self}.

The above discussion shows that there is a fundamental difference between {the gravitational control of temporal order discussed here} and other methods. Although the final state of a system undergoing some transformations in a non-classical order is independent of how the order was controlled, only when the mass controls temporal relations is the effect universal -- applying to {all} events in some space-time region.  Thus, only in the gravitational case one would conclude that non-classical temporal order indicates non-classicality of space-time.

\subsubsection*{Supplementary Note 5: What if it is {fundamentally not possible} to violate the Bell inequality for time order?}

Since a test of Bell's inequalities for temporal order has never been performed and would be very challenging,  one can also ask what if it is not possible even in principle to violate the corresponding Bell's inequalities, or if it is fundamentally not possible to satisfy assumptions of the theorem other than $3$?
If that were the case, a classical description of temporal order could always be given, 
e.g.~in terms of the classical variable $\lambda$ (introduced in Definition 1 in the main text) -- even in space-times originating from a quantum state of a massive body.  Moreover, the classical variable describing temporal order of events could  be used to define a classical time parameter according to which the systems evolve, even in scenarios involving macroscopic masses in quantum superposition states. Interestingly, this would imply that models forbidding spatial superpositions of large masses on the ground that it is not possible to define time evolution in the resulting space-time, such as refs.~\cite{ref:Karolyhazy1966, ref:Diosi1989, ref:Penrose1996, Stamp:2012, Penrose2014}
are redundant: Time would be compatible with a classical description (in terms of a hidden variable) even in the presence of quantum states of massive bodies.

\renewcommand\refname{References}

\end{document}